\documentclass[prl,aps,reprint,10pt,nofootinbib,preprintnumbers,superscriptaddress]{revtex4-1}
\pdfoutput=1
\usepackage{geometry}
\usepackage{graphicx}
\usepackage{color}
\usepackage{fullpage}
\usepackage[title]{appendix}
\usepackage{hyperref}
\hypersetup{colorlinks, citecolor=Blue, linkcolor=black, urlcolor=Blue}
\usepackage{setspace}
\usepackage[caption=false]{subfig}
\usepackage[usenames, dvipsnames]{xcolor}
\usepackage{tabularx}
\usepackage{empheq}
\usepackage{comment}
\usepackage{hepunits}
\usepackage{enumitem}

\definecolor{light-gray}{gray}{0.9}

\newcommand{\eps}{\epsilon}
\newcommand{\order}[1]{\mathcal{O}{(#1)}}

\newcommand{\be}{\begin{eqnarray}}
\newcommand{\ee}{\end{eqnarray}}

\newcommand{\T}{\text{T}}
\newcommand{\m}{\text{m}}
\newcommand{\beq}{\begin{equation}}
\newcommand{\eeq}{\end{equation}}

\newcommand{\tint}{t_\text{int}}
\newcommand{\gyy}{g_{a\gamma\gamma}}
\newcommand{\ghyy}{\hat{g}_{a\gamma\gamma}}
\newcommand{\Bvec}{\mathbf{B}}

\newcommand{\Evec}{\mathbf{E}}

\newcommand{\rhodm}{\rho_{_{\text{DM}}}}
\newcommand{\p}{\prime}
\newcommand{\w}{\omega}

\newcommand{\xiv}{\boldsymbol{\xi}}
\newcommand{\Sv}{{\bf S}}

\newcommand{\Eq}[1]{Eq.~(\ref{eq:#1})}
\newcommand{\Fig}[1]{Fig.~\ref{fig:#1}}

\newcommand{\ph}{\varphi}

\newcommand{\eqbox}[1]{\text{\fcolorbox{light-gray}{light-gray}{$#1$}}}

\definecolor{darkblue}{rgb}{0.0,0.0,0.6}
\definecolor{readableAB}{rgb}{0.7,0,0.7}
\definecolor{colorRTD}{rgb}{.2,.2,.7}

\begin{document}

\title{Heterodyne Broadband Detection of Axion Dark Matter}

\author{Asher~Berlin}
\affiliation{Center for Cosmology and Particle Physics, Department of Physics, New York University, New York, NY 10003, USA}
\author{Raffaele~Tito~D'Agnolo}
\affiliation{Institut de Physique Th\'eorique, Universit\'e Paris Saclay, CEA, F-91191 Gif-sur-Yvette, France}
\author{Sebastian~A.~R.~Ellis}
\author{Kevin Zhou}
\affiliation{SLAC National Accelerator Laboratory, 2575 Sand Hill Road, Menlo Park, CA 94025, USA}

\begin{abstract}
We propose a new broadband search strategy for ultralight axion dark matter that interacts with electromagnetism. An oscillating axion field induces transitions between two quasi-degenerate resonant modes of a superconducting cavity. In two broadband runs optimized for high and low masses, this setup can probe unexplored parameter space for axion-like particles covering fifteen orders of magnitude in mass, including astrophysically long-ranged fuzzy dark matter. 
\end{abstract}

\maketitle

\emph{Introduction.} --- Evidence for dark matter (DM) has been accumulating for almost ninety years~\cite{Zwicky:1933gu} and its microscopic nature remains one of the most important open questions in physics. Among the many DM candidates proposed in the literature, light pseudoscalar bosons with sub-eV masses have garnered considerable appeal since they generically appear in string compactifications~\cite{Svrcek:2006yi, Arvanitaki:2009fg, Stott:2017hvl} and have a simple and predictive cosmological history. Furthermore, in certain regions of parameter space they can solve the strong CP~\cite{Peccei:1977hh,Peccei:1977ur,Weinberg:1977ma,Wilczek:1977pj,Preskill:1982cy,Abbott:1982af} or electroweak hierarchy problem~\cite{Graham:2015cka, Fonseca:2018kqf, Banerjee:2018xmn}. In the ``fuzzy" mass limit ($m_{_\text{DM}} \sim 10^{-22} \ \eV$), light bosonic DM may also play a role in resolving long-standing tensions between observations and simulations of galactic structure~\cite{Goodman:2000tg,Hu:2000ke,Hui:2016ltb}. In this work, we present a new detection strategy for these DM candidates, which we refer to as axions. 

Axion DM generically couples to electromagnetism through the interaction
\be
\label{eq:photon}
-\mathcal{L} \supset \frac{1}{4} \, \gyy \, a\, F_{\mu \nu} \tilde{F}^{\mu \nu} \supset \frac{1}{2} \, \mathbf{J}_{\text{eff}} \cdot \mathbf{A}
\, ,
\ee
where $a$ is the axion field and $\mathbf{A}$ is the vector potential. In the presence of a background magnetic field $\mathbf{B}$, the axion sources an effective current density
\be
\label{eq:effective_current}
\mathbf{J}_{\text{eff}} \simeq \gyy \, \partial_t a \, \Bvec
\, .
\ee
The interaction of \Eq{photon} forms the basis of several experimental approaches to axion detection~\cite{Anastassopoulos:2017ftl,Hagmann:1998cb,Boutan:2018uoc,Du:2018uak,Brubaker:2016ktl,Braine:2019fqb,PhysRevLett.59.839,Wuensch:1989sa,Hagmann:1990tj, Zhong:2018rsr,Blout:2000uc, PhysRevD.88.102003, PhysRevLett.116.161101,Ouellet:2018beu,Gramolin:2020ict}. For instance, the time variation of $\mathbf{J}_\text{eff}$ produces an oscillating emf $\mathcal{E} \propto \partial_t J_{\text{eff}}$, which may be used to drive a resonant detector~\cite{Sikivie:1983ip,Sikivie:1985yu}. Such experiments exploit the coherence properties of the axion DM field, which we model as a classical Gaussian random field within the galaxy, with an average local density $\rhodm \simeq 0.4 \ \GeV / \cm^3$ and oscillating with angular frequency approximately equal to the axion mass $m_a$. Velocity dispersion from virialization within the galaxy leads to a spectral broadening of the axion, with a characteristic width of $\Delta \w_a \sim m_a/Q_a$, where $Q_a \sim 10^6$.

In setups applying static magnetic fields, $\mathbf{J}_\text{eff}$ oscillates with the same frequency as the axion field. Microwave cavities that are resonantly matched to the axion field can be built for $m_a \sim \mu\eV$~\cite{Braine:2019fqb}, but for lower axion masses, the required cavity volume becomes impractically large. Resonant detection of lighter axions is possible in static-field setups if the resonant frequency and volume of the detector are independent, such as for lumped-element LC circuits~\cite{Sikivie:2013laa,Kahn:2016aff,Chaudhuri:2019ntz}. However, their sensitivity to low mass axions is suppressed by $\partial_t J_{\text{eff}} \propto m_a$.

\begin{figure*}[t]
\includegraphics[width=1.55\columnwidth]{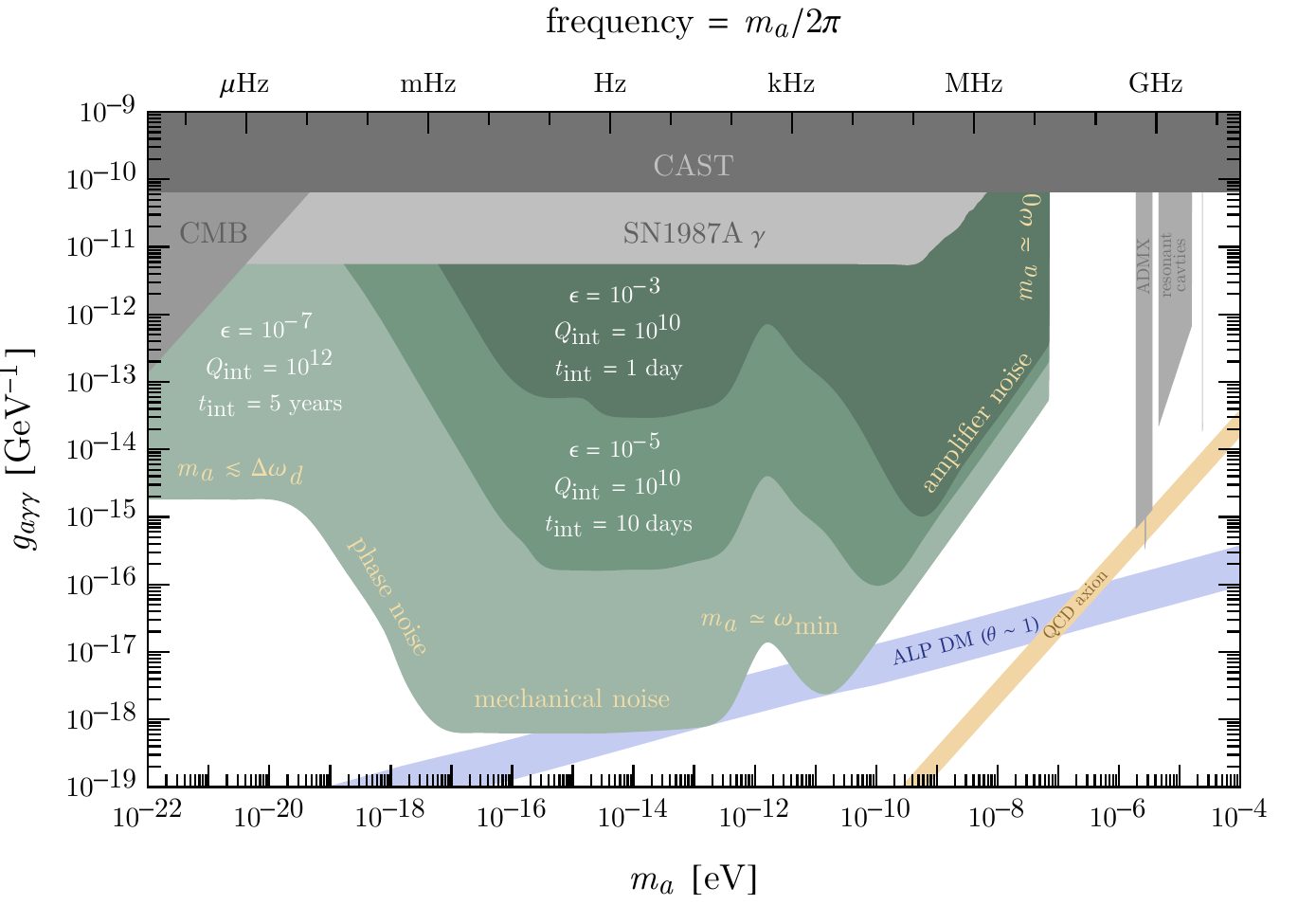}
\caption{In shaded green, the projected 90\% C.L. reach of our setup to axion dark matter for several values of leakage noise suppression $\eps$, intrinsic quality factor $Q_\text{int}$, and integration time $\tint$. We assume pump and signal mode frequencies $\w_0 = \w_1 = 100 \ \MHz$, a cavity volume $V_\text{cav} = \m^3$, a magnetic field strength $B_0 = 0.2 \ \T$, a mode overlap form factor $\eta_a = 1$, a drive oscillator width $\Delta \w_d = 0.1 \ \mHz$, and an attenuated RMS cavity wall displacement $q_\text{rms} = 0.1 \ \nm$. Further variations are shown in \Fig{reachvary}. Shown in gray are regions excluded by CAST, cavity haloscopes, measurements of the CMB, and observations of SN1987A~\cite{Anastassopoulos:2017ftl,Boutan:2018uoc,Du:2018uak, PhysRevD.69.011101, Brubaker:2016ktl, PhysRevLett.59.839,Wuensch:1989sa,Hagmann:1990tj, McAllister:2017lkb,Fedderke:2019ajk,Raffelt:1996wa,Payez:2014xsa,Zhong:2018rsr}. The orange band denotes parameter space motivated by the strong CP problem. Along the blue band, axions are produced through the misalignment mechanism at a level consistent with the observed dark matter energy density, assuming a temperature independent mass and an $\order{1}$ initial misalignment angle (see Ref.~\cite{Blinov:2019rhb} for a recent discussion), where we have assumed a symmetry breaking scale $f_a$ given by $\gyy = \alpha_\text{em} / (2 \pi f_a)$. For larger couplings above the blue band, axions produced in the same way would instead make up a subcomponent of dark matter, $\rho_a \lesssim \rhodm$. However, since $J_{\text{eff}} \propto \gyy \sqrt{\rho_a} \propto \gyy f_a$ is independent of $\gyy \propto 1/f_a$, our setup is equally sensitive to such subcomponents.}
\label{fig:reach}
\end{figure*}

Recently, we have proposed a new approach for axion DM detection, which uses frequency conversion to retain the advantages of resonant cavities while avoiding this suppression at low masses~\cite{Berlin:2019ahk} (see also Refs.~\cite{Sikivie:2010fa,Bogorad:2019pbu,Lasenby:2019prg}).\footnote{Resonant and broadband heterodyne setups based on optical interferometry have previously been proposed, but their sensitivity is limited by laser shot noise~\cite{DeRocco:2018jwe,Obata:2018vvr,Liu:2018icu}.} A cavity is prepared by driving a ``pump mode'' with frequency $\w_0 \sim \GHz$, so that the axion can resonantly drive power into a ``signal mode'' of nearly degenerate frequency $\w_1 \simeq \w_0 \pm m_a$ and distinct spatial geometry. A scan over possible axion masses is performed by slightly perturbing the cavity geometry, thereby modulating the frequency splitting $\w_1 - \w_0$. Compared to a static-field LC circuit of comparable volume and noise, the signal-to-noise ratio of this ``heterodyne'' approach is parametrically enhanced by $\w_1/m_a$. It also benefits from the very large intrinsic quality factors $Q_\text{int} \gtrsim 2 \times 10^{11}$ achievable in superconducting radio frequency (SRF) cavities~\cite{Romanenko:2014yaa,Posen:2018bjn}, which far exceed the quality factors achievable in static-field detectors targeting small axion masses.

In this work, we consider a broadband search where the signal and pump modes are fixed to be degenerate within their bandwidth, the feasibility of which has recently been demonstrated by the DarkSRF collaboration~\cite{DarkSRF}. For the lowest axion masses, $m_a \lesssim \w_0 / Q_\text{int} \sim 10^{-17} \ \eV$, the signal power is resonantly enhanced. For higher axion masses, the signal is off-resonance, but so are the dominant sources of noise in the cavity, thereby allowing this setup to explore new parameter space for axions as heavy as $m_a \sim 10^{-7} \ \eV$, as shown in \Fig{reach}. This broadband approach is thus sensitive to a wide range of axion masses without the need to scan over frequency splittings. It is also the first approach that could directly detect electromagnetically-coupled axion DM at the lowest viable DM masses $m_a \sim 10^{-22} \ \eV$, which correspond to a de Broglie wavelength the size of dwarf galaxies and a coherence time ten times longer than recorded human history. 

\begin{figure*}[t]
\includegraphics[width=1\columnwidth]{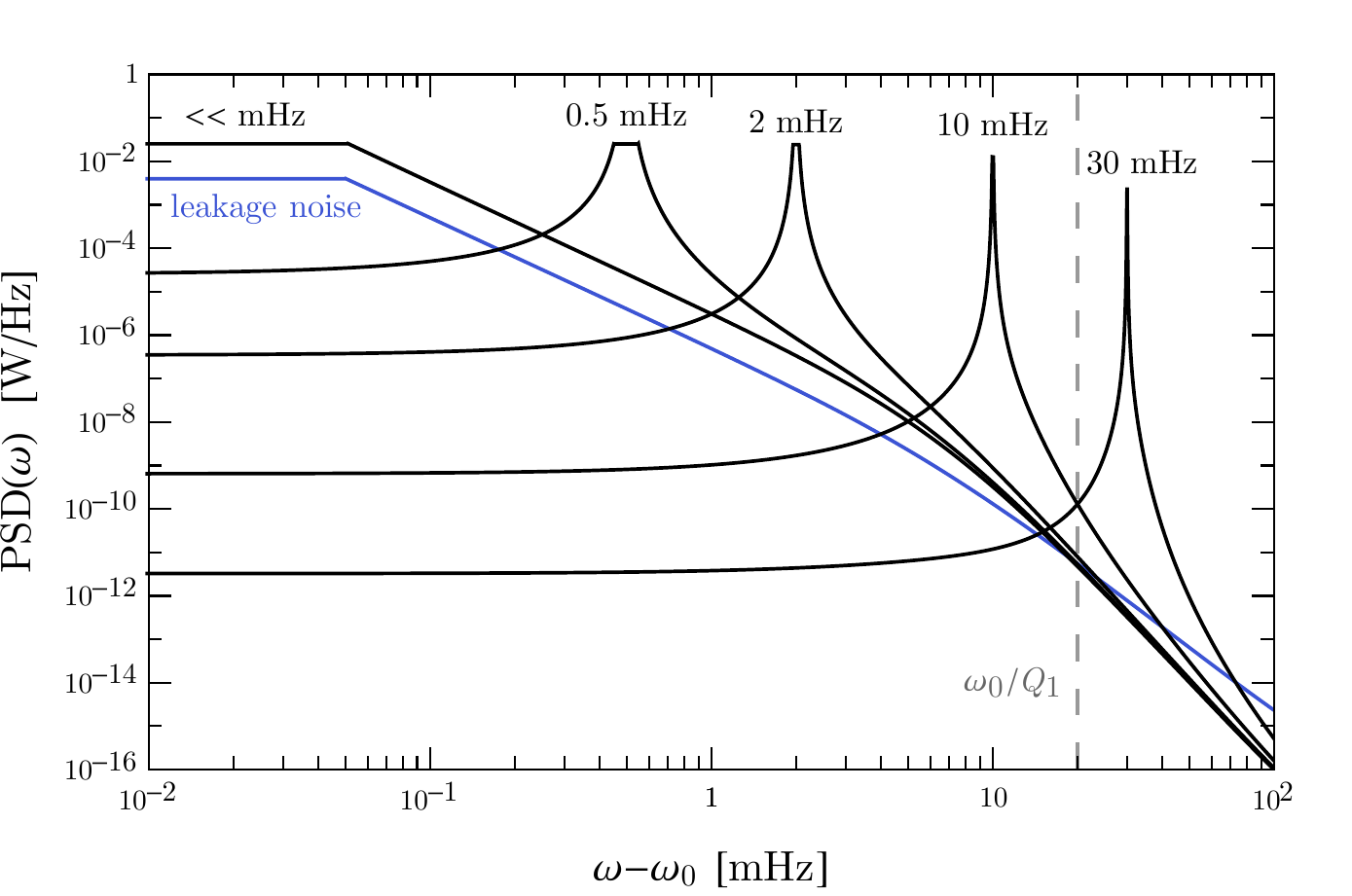}
\includegraphics[width=1\columnwidth]{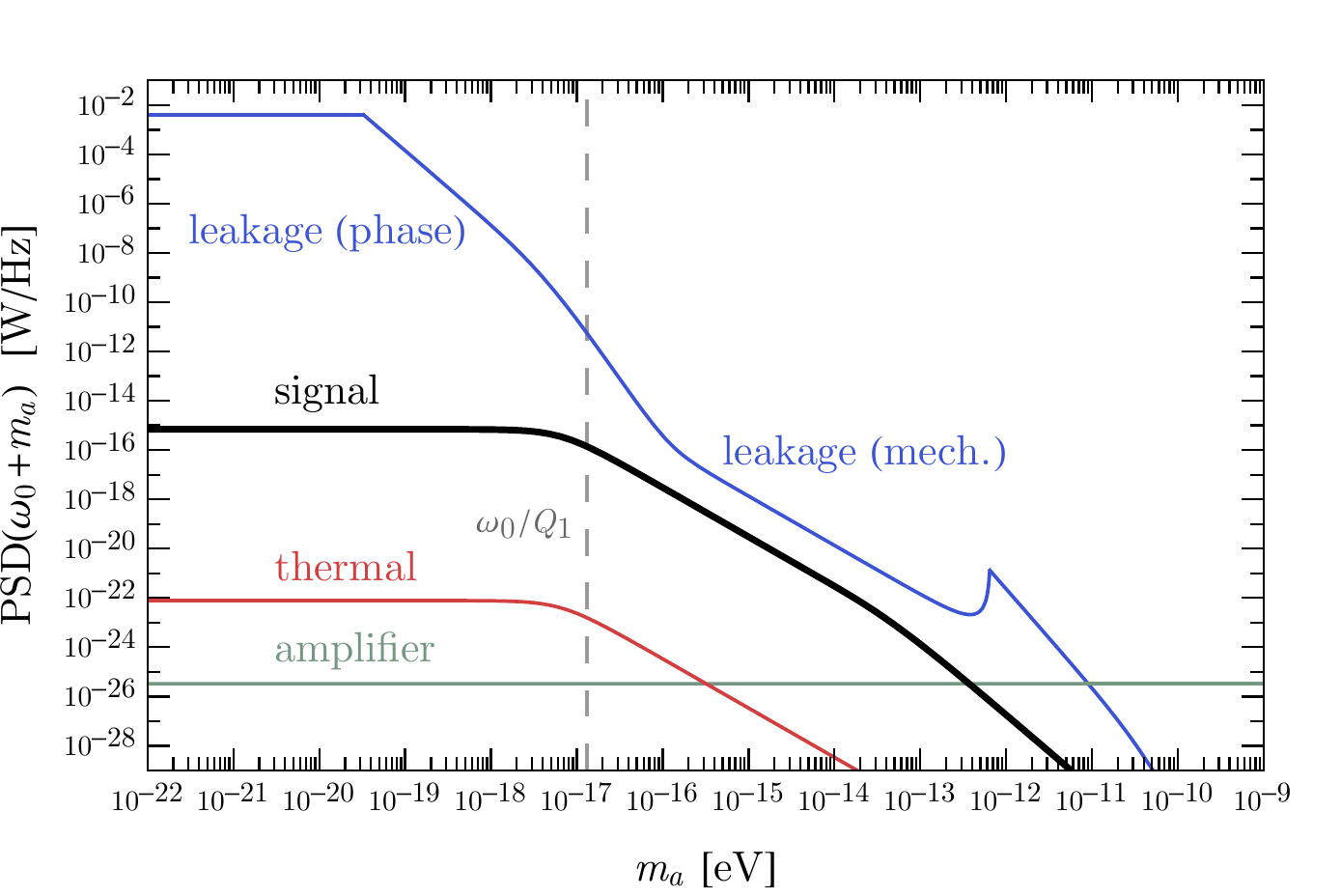}
\caption{(Left) The signal (black) and leakage noise (blue) PSDs as a function of $\w - \w_0$, for a fixed value of $\gyy$ and various choices of the axion mass (black labels), at critical coupling. The signal PSD peaks at $\w_\text{sig}$ and the vertical dashed line denotes the bandwidth $\Delta \w_r$ of the signal mode. The parameters are those of the second-lowest curve in \Fig{reach}. (Right) As in the left panel, but now showing the PSDs evaluated at $\w_{\text{sig}}$ as a function of the axion mass $m_a$. Note that we have selected a value of $\gyy$ corresponding to $\text{SNR} \ll 1$. This is done to allow the reader to more easily compare the slopes of signal and noise. We show leakage noise from the oscillator (blue), thermal fluctuations of the electromagnetic field in a cavity cooled to $1.8 \ \text{K}$ (red), and quantum-limited amplifier noise (green). Mechanical vibrations dominate the leakage noise for high $m_a$. 
}
\label{fig:PSD}
\end{figure*}

\emph{Detection Strategy.} --- Our setup involves preparing an SRF cavity by driving a loading waveguide, predominantly coupled to the pump mode, with an external oscillator at frequency $\w_0$. In the presence of axion DM, the pump mode magnetic field $\Bvec_0$ sources an effective current\footnote{The signal survives at low axion masses because in this limit $\partial_t J_{\text{eff}} \simeq \gyy \, \partial_t a \, \partial_t^2 B \propto m_a \, a \propto \sqrt{\rhodm}$ is independent of $m_a$ for a fixed axion energy density. For a fixed axion field \textit{amplitude}, $\partial_t J_\text{eff} \to 0$ as $m_a \to 0$, as required from general principles.} as in \Eq{effective_current} that oscillates at frequency
\be
\label{eq:wsig}
\w_\text{sig} \simeq \w_0 \pm m_a
\, .
\ee
Since this current is parallel to $\Bvec_0$, it drives power into the signal mode with strength parametrized by the form factor
\be
\label{eq:formfactor}
\eta_a = \frac{|\int_{V_\text{cav}} \Evec_1^* (x) \cdot \Bvec_0 (x)|}{\big( \int_{V_\text{cav}} |\Evec_1 (x)|^2 ~ \int_{V_\text{cav}} |\Bvec_0 (x)|^2 \big)^{1/2}} \, \leq 1
~,
\ee
where $\Evec_1$ is the signal mode electric field and $V_{\text{cav}}$ is the volume of the cavity. As a concrete example, $\eta_a \sim \order{1}$ for the $\text{TE}_{011}$ and $\text{TM}_{020}$ modes of a cylindrical cavity, which are degenerate in frequency for a length-to-radius ratio of $L / R \simeq 0.8$~\cite{Berlin:2019ahk,Lasenby:2019prg}. The signal is extracted through a readout waveguide, predominantly coupled to the signal mode. The frequency $\w_0$ of the pump mode and $\w_1$ of the signal mode are held fixed and taken to be degenerate within the signal mode bandwidth.

When the sensitivities of a broadband and scanning approach overlap, the latter is stronger with a similar cavity~\cite{Berlin:2019ahk}, as expected on general grounds~\cite{Chaudhuri:2018rqn}. The two approaches have the same sensitivity only when $m_a$ is smaller than the resonator bandwidth and the broadband setup functions as a resonant experiment. However, a broadband setup is simpler to operate due to its fixed geometry, and could be used as a stepping stone towards a scanning one. Moreover, it can probe a wide range of parameter space in a short integration time. 

\emph{Overview of Signal and Noise.} --- The frequency spread $\Delta \w_{\rm sig}$ of $\mathbf{J}_{\text{eff}}$ (and hence of our signal) depends on the width $\Delta \w_a$ of the axion field and the width $\Delta \w_d$ of the oscillator driving the pump mode, $\Delta \w_{\rm sig}\sim \max ( \Delta \w_a , \Delta \w_d )$. For concreteness, we take the power spectral density (PSD) of the central peak of the oscillator to be flat with a width $\Delta \w_d \simeq 0.1 \ \mHz$, comparable to a commercially available oscillator~\cite{datasheet}. This is narrower than the signal mode width $\Delta \w_r = \w_0 / Q_1$ for all parameters we consider. Since it can be beneficial to overcouple the readout, the loaded quality factor $Q_1$ of the signal mode can be much lower than the intrinsic quality factor $Q_{\text{int}}$, though the pump mode quality factor $Q_0$ remains comparable to $Q_{\text{int}}$.

The average signal power delivered to the cavity is
\be
\label{eq:Psig1}
P_\text{sig} \sim \frac{(\gyy \, \eta_a \, B_0)^2 \, \rhodm V_\text{cav}}{\text{max} \left( \Delta \w_r , \Delta \w_a \right)} \, \min \left[ 1,\, \Big(\frac{\Delta \w_r }{m_a}\Big)^2 \, \right]~,
\ee
where $B_0$ is the characteristic amplitude of the pump mode magnetic field, defined in \Eq{Bfield}. The final factor in \Eq{Psig1} accounts for the suppression that occurs when the axion drives the signal mode off-resonance ($m_a \gtrsim \Delta \w_r$). Given the signal power and noise PSD $S_n(\w)$ (examples of which are shown in \Fig{PSD}), the reach is determined by the signal-to-noise ratio~\cite{Dicke:1946glx}
\be
\label{eq:SNR0}
\text{SNR} \sim \frac{P_\text{sig}}{S_n (\w_\text{sig})} ~  \sqrt{\frac{\tint}{\Delta \w_\text{sig}}}
~,
\ee
where $\tint$ is the total integration time. \Eq{SNR0} is valid provided that $\tint \gtrsim 1/\Delta \omega_{\text{sig}}$, which holds for all parameters we consider. A detailed derivation of the signal power and of the test statistic that \Eq{SNR0} approximates is given in the Supplemental Material. 

For most of the axion masses we consider, the dominant noise source is power in the oscillator or pump mode ``leaking'' into the readout waveguide. For instance, geometric imperfections can lead to small cross-couplings $\eps \ll 1$ between the loading architecture and signal mode (and similarly between the readout and pump mode), resulting in leakage noise power proportional to $\eps^2$. Leakage noise was previously encountered in the gravitational wave experiment MAGO, which looked for transitions between nearly degenerate symmetric and antisymmetric mode combinations of two identical SRF cavities coupled by a small tunable aperture~\cite{Bernard:2001kp,Bernard:2002ci,Ballantini:2005am}. The collaboration achieved a noise suppression of $\eps \sim 10^{-7}$ using magic-tees and a variable phase shifter coupled to an active feedback loop~\cite{Bernard:2000pz}. Our setup benefits from the fact that the two modes can be chosen to be locally orthogonal, $\Evec_0 \cdot \Evec_1 = \Bvec_0 \cdot \Bvec_1 = 0$, with distinct spatial profiles. This could allow for further noise suppression by, e.g., loading/reading out the pump/signal mode near a node of the other mode~\cite{Lasenby:2019prg}, or by correlating readout measurements across multiple regions of the cavity. In the following, we conservatively consider $\epsilon \geq 10^{-7}$.

As shown in the right panel of \Fig{PSD}, leakage noise is largest when $m_a \lesssim \Delta \w_d$, while for higher axion masses it falls off according to the tail of the pump mode PSD, which is determined by oscillator ``phase noise" and mechanical vibrations of the cavity~\cite{Berlin:2019ahk}. For the highest axion masses we consider, readout amplifier noise dominates. This explains the main qualitative features of \Fig{reach}. Since slightly different setups are optimal in each mass regime (with the exact crossover points depending on the experimental parameters), we organize the following discussion by axion mass.

\emph{Low mass axions, $m_a \lesssim \Delta \w_d$.} --- When the axion mass is smaller than the oscillator width ($m_a \lesssim 10^{-19} \ \eV$), the signal overlaps in frequency with the central peak of the oscillator. Both the signal and noise are spread over a bandwidth $\Delta \w_\text{sig} \sim \Delta \w_d$, giving a leakage noise PSD of
\be
S_\text{leak} (\w_\text{sig}) \sim \eps^2 \, P_\text{in} / \Delta \w_d
~,
\ee
where $P_\text{in} \sim (\w_0 / Q_\text{int}) \, B_0^2 \, V_\text{cav}$ is the power stored in the cavity. This leads to an SNR of
\be
\label{eq:SNRlow}
\text{SNR} \sim \rhodm \, \left( \frac{\gyy \, Q_\text{int}}{\w_0 \, \eps} \right)^2 \sqrt{\tint \, \Delta \w_d} 
\ee
and hence a reach $\gyy \propto \epsilon/Q_{\text{int}}$, independent of $m_a$. We have assumed the readout waveguide is critically coupled to the signal mode ($Q_1 = Q_\text{int}/2$), which maximizes the sensitivity. Since leakage noise is the main noise source, parameters such as $B_0$, $V_\text{cav}$, and the cavity temperature do not directly affect the sensitivity. 

In contrast to precision interferometric experiments, external sources of low frequency noise, such as ground vibrations or the cooling apparatus, do not appreciably affect our reach. Relative displacements of the cavity walls are suppressed by the rigidity of the cavity and further controlled by actively monitoring the mode frequencies and cross-coupling $\epsilon$. We conservatively estimate the effect of such vibrational noise to be many orders of magnitude below leakage noise in this mass range. These points are discussed further in the Supplemental Material.

Although the signal and noise overlap in frequency, they are not indistinguishable. Other than their distinct spatial profiles and spectral tails, there are two other effects to consider. Since $m_a \lesssim \Delta \w_r$ in this regime, the axion field oscillates less than once per ring-up time of the cavity. Hence, the instantaneous signal power tracks the oscillations of $(\partial_t a)^2$, with angular frequency $2 m_a$. Furthermore, $J_\text{eff} \propto B_0$ drives the signal mode on resonance, leading to a signal mode magnetic field $\pi/2$ out of phase with leakage noise. More generally, fluctuations in leakage noise due to fluctuations in the pump mode field can be monitored and ideally subtracted out. Thus, the parameter $\epsilon$ in \Eq{SNRlow} should be regarded as including the ability to distinguish between signal and leakage noise using these additional handles. As this depends on technical details of the experimental setup, we do not attempt to estimate it here, and instead set our lowest $\epsilon$ value in \Fig{reach} to that achieved by MAGO ($\epsilon \sim 10^{-7}$).

In this regime, the axion is monochromatic up to the experimental resolution, $\Delta \w_a \lesssim 1/ \tint$, which introduces an additional subtlety. If the axion is modeled as a Gaussian random field, then its amplitude varies by an $\order{1}$ amount over the axion coherence time $\tau_a \sim 1/\Delta \w_a$. Since the average signal power $P_{\text{sig}} \propto \rhodm \propto \langle a^2 \rangle$ only applies when averaging over many coherence times, the sensitivity here is weakened due to the possibility of a downward fluctuation in the axion amplitude which lasts for the duration of the experiment. We treat this effect in the Supplemental Material with frequentist statistics (see also Refs.~\cite{Foster:2017hbq, Centers:2019dyn}). For an integration time $\tint = 1 \ \text{day}$ or $5 \ \text{years}$, it suppresses the reach in $\gyy$ by approximately a factor of $2$ for $m_a \lesssim 10^{-15}\, \eV$ or $m_a \lesssim 10^{-18} \ \eV$, respectively. 

\emph{High mass axions, $m_a \gtrsim \kHz$.} --- Here, leakage and thermal noise are negligible due to the off-resonance suppression $(\Delta \w_r / m_a)^2$. Since the axion is wider than the oscillator, the signal width is $\Delta \w_\text{sig} \sim \Delta \w_a$, and amplifier noise dominates as in static broadband axion searches in this mass range~\cite{Kahn:2016aff,Gramolin:2020ict}, such that
\be
\text{SNR} \sim \rhodm V_\text{cav} \, \frac{\Delta \w_r}{S_\text{amp} (\w_\text{sig})} \Big( \frac{\gyy \, B_0}{m_a} \Big)^2 \sqrt{\frac{\tint}{\Delta \w_a}}
~ .
\ee
Here, we assume a quantum-limited amplifier, $S_{\text{amp}}(\w) \sim \hbar \w$. This is only feasible for high mass axions; for lower axion masses the amplifier would be saturated by leakage noise~\cite{liu2017josephson}. We cut off the reach in \Fig{reach} at $m_a \simeq \w_0$, above which higher harmonics of the cavity must be considered~\cite{Sikivie:2010fa}, as well as potential nonlinear response of the cavity walls~\cite{Eriksson:2004cz}. 

The reach scales as $\gyy \propto m_a^{5/4} / \Delta \w_r^{1/2}$, assuming $\Delta \w_r \gtrsim \Delta \w_a$. Thus, a lower $Q_1$ is beneficial when amplifier noise dominates, since it reduces the suppression of the signal. This can be achieved by overcoupling the signal mode to the readout; we set $Q_1 \sim 10^5$, which is a typical loaded quality factor of SRF cavities in accelerators~\cite{padamsee2009rf}.


\emph{Intermediate mass axions, $\Delta \w_d \lesssim m_a \lesssim \kHz$.} --- For the bulk of the parameter space shown in \Fig{reach}, the reach is dictated by the high frequency tail of the leakage noise. In most of this range, the oscillator is wider than the axion, so the signal width is $\Delta \w_\text{sig} \sim \Delta \w_d$. 

In the lower end of this mass range, the main contribution to the leakage noise tail is from oscillator phase noise~\cite{Berlin:2019ahk}, which for $m_a \gtrsim \Delta \w_r$ is of the form
\be
S_\text{leak} (\w_\text{sig}) \sim \eps^2 \, P_\text{in} \, \Big( \frac{\Delta \w_r}{m_a} \Big)^2 \, S_\ph (m_a)
~,
\ee
where the phase noise PSD $S_\ph (\w)$ is parametrized by~\cite{rubiola2009phase}
\be
S_\ph(\w) = \sum_{n=0}^3 c_n \, \w^{-n}
~,
\ee
and the $c_n$ are fit to a commercially available oscillator~\cite{datasheet}. For $m_a$ slightly higher than $\Delta \w_d$, the cubic term in $S_\ph(\w)$ dominates, resulting in $S_\text{leak} (\w_\text{sig}) \propto 1/m_a^5$ and a rapid improvement in the reach at higher axion masses. 

In the upper end of this mass range, the main noise contribution instead arises from displacements of the cavity walls, where mechanical vibrations at frequency $m_a$ contribute to pump mode power at $\w_\text{sig}$~\cite{Berlin:2019ahk}.
On the basis of previous measurements in a MAGO prototype~\cite{Bernard:2001kp}, we take the external mechanical force PSD to be spectrally flat, and the mechanical modes to have a quality factor $Q_m \sim 10^3$. As described in the Supplemental Material, the contribution of the lowest-lying mechanical resonance at $\w_{\text{min}} \sim \text{kHz}$ dominates for $m_a \lesssim \w_\text{min}$, such that
\be
\label{eq:mechleak}
S_\text{leak} (\w_\text{sig}) \sim \eps^2 \, P_\text{in} \, \Big( \frac{\Delta \w_r}{m_a} \Big)^2 \,  \frac{\delta^2 \, Q_\text{int}^2}{\w_\text{min} \, Q_m}
~,
\ee
where $\delta \ll 1$ is the fractional displacement of the cavity walls. For $\Delta \w_a \lesssim \Delta \w_r \lesssim m_a$, $P_\text{sig} \propto 1/m_a^2$, and thus the sensitivity in this region is independent of the axion mass. For frequencies above $\w_\text{min}$, we assume that a forest of evenly spaced mechanical modes exists. To estimate $\delta$, we note that the DarkSRF collaboration has recently demonstrated the ability to control the resonant frequency of a driven cavity to one part in $Q_\text{int} \gtrsim 10^{10}$, corresponding to sub-nm displacements of the cavity walls~\cite{fnalex,DarkSRF}. This has been demonstrated on minute timescales, and a near-future run is expected to prolong this to $\order{1} \text{ week}$. Thus, we fix the typical RMS cavity wall displacement to $q_\text{rms} = 0.1 \ \text{nm}$, corresponding to $\delta \sim 10^{-10}$ for a meter-sized cavity. This is larger than the displacement due to environmental seismic noise~\cite{Saulson:2017jlf}, reflecting the expectation that vibrations will primarily arise from the apparatus itself (e.g., the helium pump). 

Deformations of the cavity walls can also directly transfer power between the pump and signal modes. This ``mode mixing'' is parametrized by a dimensionless mechanical form factor $\eta_{\text{mix}}$, with $S_{\text{mix}} \sim (\eta_{\text{mix}} / \epsilon)^2 \, S_{\text{leak}}$. The form factor $\eta_{\text{mix}}$ vanishes for a perfectly cylindrical cavity, which implies its value is set by cavity deformations~\cite{Meidlinger,Bernard:2002ci}. Since $\epsilon$ parametrizes the precision to which we can control slow deformations of the cavity and waveguide geometry, we expect $\eta_{\text{mix}} \sim \epsilon$, such that mode mixing is at most comparable to mechanical leakage noise.

\emph{Run Optimization.} --- Overcoupling improves the sensitivity when amplifier noise dominates, but also shrinks the mass range where this is the case. The vast majority of the reach of \Fig{reach} can be attained as the envelope of two distinct experimental runs: (1) a critically coupled run targeting low masses and (2) a strongly overcoupled run with a quantum-limited amplifier targeting high masses. For the lowest curve in \Fig{reach}, the full sensitivity to intermediate masses requires an additional run with less overcoupling. Only the critically coupled run benefits from a very high $Q_{\text{int}}$, while only the overcoupled runs require high $B_0$. We do not consider $m_a \lesssim 1/\tint$, in which case the reach is suppressed by the unknown instantaneous phase of the axion; for $m_a \gtrsim 1/\tint$, the sensitivity scales weakly with the integration time as $\gyy \propto 1/\tint^{1/4}$. This requires $\tint \gtrsim \text{year}$ for $m_a \sim 10^{-22} \ \eV$, which corresponds to the lowest curve in \Fig{reach}. We note that comparable parameter space can also be explored with a few much shorter runs spaced out over the course of a year, such that the axion phase is fixed within each individual run but varies between adjacent runs.  Unexplored parameter space spanning decades of axion mass could therefore be probed with typical SRF cavity quality factors $Q_{\text{int}} \sim 10^{10}$ and as little as a day of data-taking.

\begin{figure*}[t]
\includegraphics[width=2.17\columnwidth]{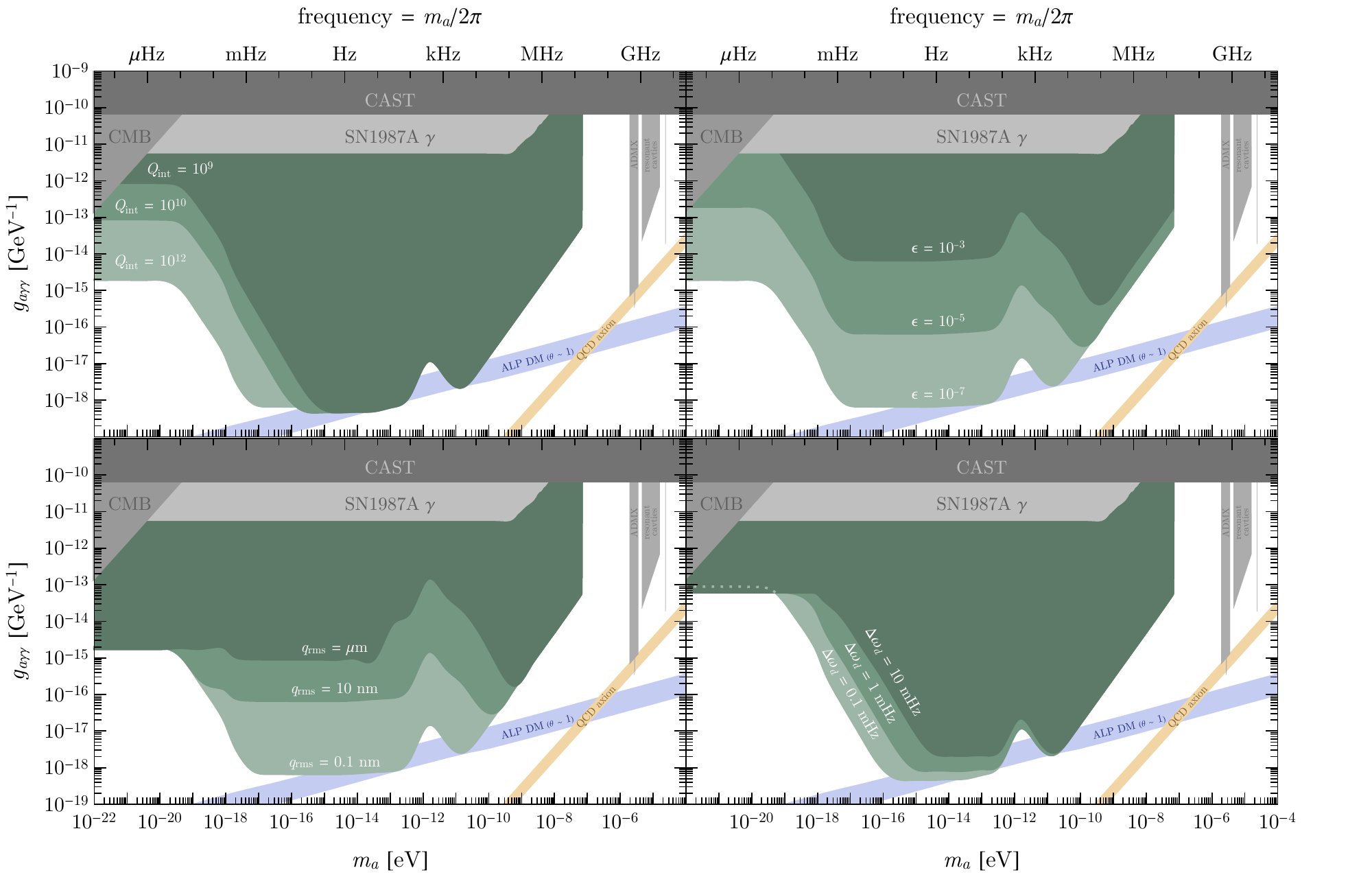}
\caption{The projected 90\% C.L. reach of our setup for a wider range of experimental parameters. The benchmark parameters are those of the lowest curve of \Fig{reach}, except for the lower-right panel, where we take $Q_\text{int} = 10^{10}$. The green dotted line in this panel shows the reach for $\Delta \w_d = 0.1 \ \mHz$, demonstrating the small effect of $\Delta \w_d$ on the sensitivity to small axion masses.}
\label{fig:reachvary}
\end{figure*}

\emph{Variations of Experimental Parameters.} --- To demonstrate the robustness of the approach, in \Fig{reachvary} we show the expected reach for experimental parameters which are orders of magnitude worse than the state of the art. 

\begin{itemize}[leftmargin=*] 
\item In the upper-left panel, we vary the intrinsic quality factor $Q_{\text{int}}$. Lowering $Q_\text{int}$ only has an adverse effect at the lowest axion masses; for higher axion masses there is no effect because the signal mode is taken to be strongly overcoupled in this mass range. 
\item In the upper-right panel, we vary the leakage noise suppression $\epsilon$. Even for $\epsilon = 10^{-3}$ (four orders of magnitude above that measured by MAGO), corresponding to straightforward millimeter-level control of the cavity geometry, substantial new parameter space can be covered. 
\item In the lower-left panel, we increase the attenuated displacement of the cavity walls by four orders of magnitude. Increasing $q_\text{rms}$ lowers the reach at intermediate masses, where mechanical noise dominates, but leaves the sensitivity to other axion masses unchanged. 
\item In the lower-right panel we increase $\Delta \w_d$. As discussed in the Supplemental Material, this mimics the effect of increased low frequency noise in the form of slow drifts of the resonant frequencies over the range $\w_0 \pm \Delta \w_d$. Since we have assumed that the pump and signal modes can be held degenerate (equivalent to $\Delta \w_d \lesssim \Delta \w_r$), we have decreased the quality factor to $Q_{\text{int}} = 10^{10}$ for this panel. Increasing $\Delta \w_d$ has little effect at low axion masses because the signal and noise already overlap completely in frequency. However, for intermediate masses, larger $\Delta \w_d$ decreases the sensitivity since it broadens the signal compared to the dominant noise source. At higher axion masses, there is no effect because $\Delta \w_a \gtrsim \Delta \w_d$.
\end{itemize}
%

\emph{Discussion.} --- We have proposed a heterodyne approach to search for ultralight axion dark matter through its coupling to electromagnetism, which applies recent developments in the manufacturing and control of SRF cavities. Due to the decreasing signal power and increasing strength of readout noise at low frequencies, traditional static-field haloscopes have limited reach to axions lighter than a $\kHz \sim 10^{-12} \ \eV$~\cite{Ouellet:2018beu,Gramolin:2020ict}. In contrast, our setup is sensitive to much lighter axions, including the entire allowed mass range for fuzzy dark matter, $m_a \geq \order{10^{-21}} \ \eV$~\cite{Armengaud:2017nkf,Irsic:2017yje,Nori:2018pka,Garzilli:2019qki,Marsh:2018zyw,Schutz:2020jox}, thereby complementing ultralight axion searches that use non-electromagnetic couplings~\cite{graham2018spin,wu2019search,terrano2019constraints,abel2017search}. It is also sensitive to axions as heavy as $10^{-7} \ \eV$, including those motivated by string theory~\cite{Halverson:2019cmy} and the misalignment mechanism. 

Our projections rely on noise estimates that are anchored to experimental findings, such as those obtained fifteen years ago by the MAGO collaboration~\cite{Bernard:2001kp,Bernard:2002ci,Ballantini:2005am}. More recently, there has been renewed interest in the SRF community to apply their technological advances to new physics searches, leading to the recent results of the DarkSRF collaboration~\cite{DarkSRF} that show the feasibility of our proposed approach. The promising sensitivity of SRF cavities to weakly coupled physics, demonstrated in this work, motivates in situ measurements of mode mixing and leakage noise, in order to further investigate the potential of these ideas. Future developments, some of which are already envisioned by the DarkSRF collaboration, can further extend our reach, improving the capacity to probe some of the most motivated dark matter candidates. 

\begin{acknowledgments}
\emph{Acknowledgments.} --- We thank Saptarshi Chaudhuri, Peter Graham, Roni Harnik, Robert Lasenby, Christopher Nantista, Jeffrey Neilson, Philip Schuster, Sami Tantawi, and Natalia Toro for valuable discussions. AB is supported by the James Arthur Fellowship. SARE is supported by the U.S.\ Department of Energy under Contract No.\ DE-AC02-76SF00515 and by the Swiss National Science Foundation, SNF project number P400P2$\_$186678. KZ is supported by the NSF GRFP under grant DGE-1656518.
\end{acknowledgments}


\bibliographystyle{utphys}
\bibliography{AxionSRF}

\providecommand{\href}[2]{#2}\begingroup\raggedright\begin{thebibliography}{10}

\bibitem{Zwicky:1933gu}
F.~Zwicky, ``{Die Rotverschiebung von extragalaktischen Nebeln},''
  \href{http://dx.doi.org/10.1007/s10714-008-0707-4}{{\em Helv. Phys. Acta}
  {\bf 6} (1933)  110--127}.

\bibitem{Svrcek:2006yi}
P.~Svrcek and E.~Witten, ``{Axions In String Theory},''
  \href{http://dx.doi.org/10.1088/1126-6708/2006/06/051}{{\em JHEP} {\bf 06}
  (2006)  051},
\href{http://arxiv.org/abs/hep-th/0605206}{{\tt arXiv:hep-th/0605206
  [hep-th]}}.

\bibitem{Arvanitaki:2009fg}
A.~Arvanitaki, S.~Dimopoulos, S.~Dubovsky, N.~Kaloper, and J.~March-Russell,
  ``{String Axiverse},''
  \href{http://dx.doi.org/10.1103/PhysRevD.81.123530}{{\em Phys. Rev.} {\bf
  D81} (2010)  123530},
\href{http://arxiv.org/abs/0905.4720}{{\tt arXiv:0905.4720 [hep-th]}}.

\bibitem{Stott:2017hvl}
M.~J. Stott, D.~J.~E. Marsh, C.~Pongkitivanichkul, L.~C. Price, and B.~S.
  Acharya, ``{Spectrum of the axion dark sector},''
  \href{http://dx.doi.org/10.1103/PhysRevD.96.083510}{{\em Phys. Rev.} {\bf
  D96} (2017) no.~8, 083510},
\href{http://arxiv.org/abs/1706.03236}{{\tt arXiv:1706.03236 [astro-ph.CO]}}.

\bibitem{Peccei:1977hh}
R.~D. Peccei and H.~R. Quinn, ``{CP Conservation in the Presence of
  Instantons},'' \href{http://dx.doi.org/10.1103/PhysRevLett.38.1440}{{\em
  Phys. Rev. Lett.} {\bf 38} (1977)  1440--1443}.
[,328(1977)].

\bibitem{Peccei:1977ur}
R.~D. Peccei and H.~R. Quinn, ``{Constraints Imposed by CP Conservation in the
  Presence of Instantons},''
\href{http://dx.doi.org/10.1103/PhysRevD.16.1791}{{\em Phys. Rev.} {\bf D16}
  (1977)  1791--1797}.

\bibitem{Weinberg:1977ma}
S.~Weinberg, ``{A New Light Boson?},''
\href{http://dx.doi.org/10.1103/PhysRevLett.40.223}{{\em Phys. Rev. Lett.} {\bf
  40} (1978)  223--226}.

\bibitem{Wilczek:1977pj}
F.~Wilczek, ``{Problem of Strong $P$ and $T$ Invariance in the Presence of
  Instantons},''
\href{http://dx.doi.org/10.1103/PhysRevLett.40.279}{{\em Phys. Rev. Lett.} {\bf
  40} (1978)  279--282}.

\bibitem{Preskill:1982cy}
J.~Preskill, M.~B. Wise, and F.~Wilczek, ``{Cosmology of the Invisible
  Axion},''
\href{http://dx.doi.org/10.1016/0370-2693(83)90637-8}{{\em Phys. Lett.} {\bf
  120B} (1983)  127--132}.

\bibitem{Abbott:1982af}
L.~F. Abbott and P.~Sikivie, ``{A Cosmological Bound on the Invisible Axion},''
\href{http://dx.doi.org/10.1016/0370-2693(83)90638-X}{{\em Phys. Lett.} {\bf
  B120} (1983)  133--136}.

\bibitem{Graham:2015cka}
P.~W. Graham, D.~E. Kaplan, and S.~Rajendran, ``{Cosmological Relaxation of the
  Electroweak Scale},''
  \href{http://dx.doi.org/10.1103/PhysRevLett.115.221801}{{\em Phys. Rev.
  Lett.} {\bf 115} (2015) no.~22, 221801},
  \href{http://arxiv.org/abs/1504.07551}{{\tt arXiv:1504.07551 [hep-ph]}}.

\bibitem{Fonseca:2018kqf}
N.~Fonseca and E.~Morgante, ``{Relaxion Dark Matter},''
  \href{http://dx.doi.org/10.1103/PhysRevD.100.055010}{{\em Phys. Rev. D} {\bf
  100} (2019) no.~5, 055010}, \href{http://arxiv.org/abs/1809.04534}{{\tt
  arXiv:1809.04534 [hep-ph]}}.

\bibitem{Banerjee:2018xmn}
A.~Banerjee, H.~Kim, and G.~Perez, ``{Coherent relaxion dark matter},''
  \href{http://dx.doi.org/10.1103/PhysRevD.100.115026}{{\em Phys. Rev. D} {\bf
  100} (2019) no.~11, 115026}, \href{http://arxiv.org/abs/1810.01889}{{\tt
  arXiv:1810.01889 [hep-ph]}}.

\bibitem{Goodman:2000tg}
J.~Goodman, ``{Repulsive dark matter},''
  \href{http://dx.doi.org/10.1016/S1384-1076(00)00015-4}{{\em New Astron.} {\bf
  5} (2000)  103}, \href{http://arxiv.org/abs/astro-ph/0003018}{{\tt
  arXiv:astro-ph/0003018}}.

\bibitem{Hu:2000ke}
W.~Hu, R.~Barkana, and A.~Gruzinov, ``{Cold and fuzzy dark matter},''
  \href{http://dx.doi.org/10.1103/PhysRevLett.85.1158}{{\em Phys. Rev. Lett.}
  {\bf 85} (2000)  1158--1161},
\href{http://arxiv.org/abs/astro-ph/0003365}{{\tt arXiv:astro-ph/0003365
  [astro-ph]}}.

\bibitem{Hui:2016ltb}
L.~Hui, J.~P. Ostriker, S.~Tremaine, and E.~Witten, ``{Ultralight scalars as
  cosmological dark matter},''
  \href{http://dx.doi.org/10.1103/PhysRevD.95.043541}{{\em Phys. Rev.} {\bf
  D95} (2017) no.~4, 043541},
\href{http://arxiv.org/abs/1610.08297}{{\tt arXiv:1610.08297 [astro-ph.CO]}}.

\bibitem{Anastassopoulos:2017ftl}
{\bf CAST} Collaboration, V.~Anastassopoulos {\em et al.}, ``{New CAST Limit on
  the Axion-Photon Interaction},''
  \href{http://dx.doi.org/10.1038/nphys4109}{{\em Nature Phys.} {\bf 13} (2017)
   584--590},
\href{http://arxiv.org/abs/1705.02290}{{\tt arXiv:1705.02290 [hep-ex]}}.

\bibitem{Hagmann:1998cb}
{\bf ADMX} Collaboration, C.~Hagmann {\em et al.}, ``{Results from a high
  sensitivity search for cosmic axions},''
  \href{http://dx.doi.org/10.1103/PhysRevLett.80.2043}{{\em Phys. Rev. Lett.}
  {\bf 80} (1998)  2043--2046},
  \href{http://arxiv.org/abs/astro-ph/9801286}{{\tt arXiv:astro-ph/9801286}}.

\bibitem{Boutan:2018uoc}
{\bf ADMX} Collaboration, C.~Boutan {\em et al.}, ``{Piezoelectrically Tuned
  Multimode Cavity Search for Axion Dark Matter},''
  \href{http://dx.doi.org/10.1103/PhysRevLett.121.261302}{{\em Phys. Rev.
  Lett.} {\bf 121} (2018) no.~26, 261302},
\href{http://arxiv.org/abs/1901.00920}{{\tt arXiv:1901.00920 [hep-ex]}}.

\bibitem{Du:2018uak}
{\bf ADMX} Collaboration, N.~Du {\em et al.}, ``{A Search for Invisible Axion
  Dark Matter with the Axion Dark Matter Experiment},''
  \href{http://dx.doi.org/10.1103/PhysRevLett.120.151301}{{\em Phys. Rev.
  Lett.} {\bf 120} (2018) no.~15, 151301},
\href{http://arxiv.org/abs/1804.05750}{{\tt arXiv:1804.05750 [hep-ex]}}.

\bibitem{Brubaker:2016ktl}
B.~M. Brubaker {\em et al.}, ``{First results from a microwave cavity axion
  search at 24 $\mu$eV},''
  \href{http://dx.doi.org/10.1103/PhysRevLett.118.061302}{{\em Phys. Rev.
  Lett.} {\bf 118} (2017) no.~6, 061302},
\href{http://arxiv.org/abs/1610.02580}{{\tt arXiv:1610.02580 [astro-ph.CO]}}.

\bibitem{Braine:2019fqb}
{\bf ADMX} Collaboration, T.~Braine {\em et al.}, ``{Extended Search for the
  Invisible Axion with the Axion Dark Matter Experiment},''
  \href{http://dx.doi.org/10.1103/PhysRevLett.124.101303}{{\em Phys. Rev.
  Lett.} {\bf 124} (2020) no.~10, 101303},
  \href{http://arxiv.org/abs/1910.08638}{{\tt arXiv:1910.08638 [hep-ex]}}.

\bibitem{PhysRevLett.59.839}
S.~DePanfilis, A.~C. Melissinos, B.~E. Moskowitz, J.~T. Rogers, Y.~K.
  Semertzidis, W.~U. Wuensch, H.~J. Halama, A.~G. Prodell, W.~B. Fowler, and
  F.~A. Nezrick, \href{http://dx.doi.org/10.1103/PhysRevLett.59.839}{``Limits
  on the abundance and coupling of cosmic axions at ${m}_{a}$
  \ensuremath{\mu}ev,''{\em Phys. Rev. Lett.} {\bf 59} (Aug, 1987)  839--842}.
  \url{https://link.aps.org/doi/10.1103/PhysRevLett.59.839}.

\bibitem{Wuensch:1989sa}
W.~Wuensch, S.~De~Panfilis-Wuensch, Y.~K. Semertzidis, J.~T. Rogers, A.~C.
  Melissinos, H.~J. Halama, B.~E. Moskowitz, A.~G. Prodell, W.~B. Fowler, and
  F.~A. Nezrick, ``{Results of a Laboratory Search for Cosmic Axions and Other
  Weakly Coupled Light Particles},''
\href{http://dx.doi.org/10.1103/PhysRevD.40.3153}{{\em Phys. Rev.} {\bf D40}
  (1989)  3153}.

\bibitem{Hagmann:1990tj}
C.~Hagmann, P.~Sikivie, N.~S. Sullivan, and D.~B. Tanner, ``{Results from a
  search for cosmic axions},''
\href{http://dx.doi.org/10.1103/PhysRevD.42.1297}{{\em Phys. Rev.} {\bf D42}
  (1990)  1297--1300}.

\bibitem{Zhong:2018rsr}
{\bf HAYSTAC} Collaboration, L.~Zhong {\em et al.}, ``{Results from phase 1 of
  the HAYSTAC microwave cavity axion experiment},''
  \href{http://dx.doi.org/10.1103/PhysRevD.97.092001}{{\em Phys. Rev.} {\bf
  D97} (2018) no.~9, 092001},
\href{http://arxiv.org/abs/1803.03690}{{\tt arXiv:1803.03690 [hep-ex]}}.

\bibitem{Blout:2000uc}
B.~D. Blout, E.~J. Daw, M.~P. Decowski, P.~T.~P. Ho, L.~J. Rosenberg, and D.~B.
  Yu, ``{A Radio telescope search for axions},''
  \href{http://dx.doi.org/10.1086/318310}{{\em Astrophys. J.} {\bf 546} (2001)
  825--828},
\href{http://arxiv.org/abs/astro-ph/0006310}{{\tt arXiv:astro-ph/0006310
  [astro-ph]}}.

\bibitem{PhysRevD.88.102003}
{\bf H.E.S.S.} Collaboration, A.~Abramowski {\em et al.},
  \href{http://dx.doi.org/10.1103/PhysRevD.88.102003}{``Constraints on
  axionlike particles with h.e.s.s. from the irregularity of the pks
  $2155\ensuremath{-}304$ energy spectrum,''{\em Phys. Rev. D} {\bf 88} (Nov,
  2013)  102003}. \url{https://link.aps.org/doi/10.1103/PhysRevD.88.102003}.

\bibitem{PhysRevLett.116.161101}
{\bf The Fermi-LAT} Collaboration, M.~Ajello {\em et al.},
  \href{http://dx.doi.org/10.1103/PhysRevLett.116.161101}{``Search for spectral
  irregularities due to photon--axionlike-particle oscillations with the fermi
  large area telescope,''{\em Phys. Rev. Lett.} {\bf 116} (Apr, 2016)  161101}.
  \url{https://link.aps.org/doi/10.1103/PhysRevLett.116.161101}.

\bibitem{Ouellet:2018beu}
J.~L. Ouellet {\em et al.}, ``{First Results from ABRACADABRA-10 cm: A Search
  for Sub-$\mu$eV Axion Dark Matter},''
  \href{http://dx.doi.org/10.1103/PhysRevLett.122.121802}{{\em Phys. Rev.
  Lett.} {\bf 122} (2019) no.~12, 121802},
\href{http://arxiv.org/abs/1810.12257}{{\tt arXiv:1810.12257 [hep-ex]}}.

\bibitem{Gramolin:2020ict}
A.~V. Gramolin, D.~Aybas, D.~Johnson, J.~Adam, and A.~O. Sushkov, ``{Search for
  axion-like dark matter with ferromagnets},''
  \href{http://arxiv.org/abs/2003.03348}{{\tt arXiv:2003.03348 [hep-ex]}}.

\bibitem{Sikivie:1983ip}
P.~Sikivie, ``{Experimental Tests of the Invisible Axion},''
\href{http://dx.doi.org/10.1103/PhysRevLett.51.1415,
  10.1103/PhysRevLett.52.695.2}{{\em Phys. Rev. Lett.} {\bf 51} (1983)
  1415--1417}.

\bibitem{Sikivie:1985yu}
P.~Sikivie, ``{Detection Rates for 'Invisible' Axion Searches},''
  \href{http://dx.doi.org/10.1103/PhysRevD.36.974,
  10.1103/PhysRevD.32.2988}{{\em Phys. Rev.} {\bf D32} (1985)  2988}.
[Erratum: Phys. Rev.D36,974(1987)].

\bibitem{Sikivie:2013laa}
P.~Sikivie, N.~Sullivan, and D.~B. Tanner, ``{Proposal for Axion Dark Matter
  Detection Using an LC Circuit},''
  \href{http://dx.doi.org/10.1103/PhysRevLett.112.131301}{{\em Phys. Rev.
  Lett.} {\bf 112} (2014) no.~13, 131301},
\href{http://arxiv.org/abs/1310.8545}{{\tt arXiv:1310.8545 [hep-ph]}}.

\bibitem{Kahn:2016aff}
Y.~Kahn, B.~R. Safdi, and J.~Thaler, ``{Broadband and Resonant Approaches to
  Axion Dark Matter Detection},''
  \href{http://dx.doi.org/10.1103/PhysRevLett.117.141801}{{\em Phys. Rev.
  Lett.} {\bf 117} (2016) no.~14, 141801},
\href{http://arxiv.org/abs/1602.01086}{{\tt arXiv:1602.01086 [hep-ph]}}.

\bibitem{Chaudhuri:2019ntz}
S.~Chaudhuri, K.~D. Irwin, P.~W. Graham, and J.~Mardon, ``{Optimal
  Electromagnetic Searches for Axion and Hidden-Photon Dark Matter},''
\href{http://arxiv.org/abs/1904.05806}{{\tt arXiv:1904.05806 [hep-ex]}}.

\bibitem{PhysRevD.69.011101}
S.~J. Asztalos, R.~F. Bradley, L.~Duffy, C.~Hagmann, D.~Kinion, D.~M. Moltz,
  L.~J. Rosenberg, P.~Sikivie, W.~Stoeffl, N.~S. Sullivan, D.~B. Tanner, K.~van
  Bibber, and D.~B. Yu,
  \href{http://dx.doi.org/10.1103/PhysRevD.69.011101}{``Improved rf cavity
  search for halo axions,''{\em Phys. Rev. D} {\bf 69} (Jan, 2004)  011101}.
  \url{https://link.aps.org/doi/10.1103/PhysRevD.69.011101}.

\bibitem{McAllister:2017lkb}
B.~T. McAllister, G.~Flower, E.~N. Ivanov, M.~Goryachev, J.~Bourhill, and M.~E.
  Tobar, ``{The ORGAN Experiment: An axion haloscope above 15 GHz},''
  \href{http://dx.doi.org/10.1016/j.dark.2017.09.010}{{\em Phys. Dark Univ.}
  {\bf 18} (2017)  67--72},
\href{http://arxiv.org/abs/1706.00209}{{\tt arXiv:1706.00209
  [physics.ins-det]}}.

\bibitem{Fedderke:2019ajk}
M.~A. Fedderke, P.~W. Graham, and S.~Rajendran, ``{Axion Dark Matter Detection
  with CMB Polarization},''
  \href{http://dx.doi.org/10.1103/PhysRevD.100.015040}{{\em Phys. Rev.} {\bf
  D100} (2019) no.~1, 015040},
\href{http://arxiv.org/abs/1903.02666}{{\tt arXiv:1903.02666 [astro-ph.CO]}}.

\bibitem{Raffelt:1996wa}
G.~G. Raffelt, {\em {Stars as laboratories for fundamental physics}}.
\newblock 1996.
\newblock
\url{http://wwwth.mpp.mpg.de/members/raffelt/mypapers/199613.pdf}.
\newblock

\bibitem{Payez:2014xsa}
A.~Payez, C.~Evoli, T.~Fischer, M.~Giannotti, A.~Mirizzi, and A.~Ringwald,
  ``{Revisiting the SN1987A gamma-ray limit on ultralight axion-like
  particles},'' \href{http://dx.doi.org/10.1088/1475-7516/2015/02/006}{{\em
  JCAP} {\bf 1502} (2015) no.~02, 006},
\href{http://arxiv.org/abs/1410.3747}{{\tt arXiv:1410.3747 [astro-ph.HE]}}.

\bibitem{Blinov:2019rhb}
N.~Blinov, M.~J. Dolan, P.~Draper, and J.~Kozaczuk, ``{Dark matter targets for
  axionlike particle searches},''
  \href{http://dx.doi.org/10.1103/PhysRevD.100.015049}{{\em Phys. Rev.} {\bf
  D100} (2019) no.~1, 015049},
\href{http://arxiv.org/abs/1905.06952}{{\tt arXiv:1905.06952 [hep-ph]}}.

\bibitem{DeRocco:2018jwe}
W.~DeRocco and A.~Hook, ``{Axion interferometry},''
  \href{http://dx.doi.org/10.1103/PhysRevD.98.035021}{{\em Phys. Rev.} {\bf
  D98} (2018) no.~3, 035021},
\href{http://arxiv.org/abs/1802.07273}{{\tt arXiv:1802.07273 [hep-ph]}}.

\bibitem{Obata:2018vvr}
I.~Obata, T.~Fujita, and Y.~Michimura, ``{Optical Ring Cavity Search for Axion
  Dark Matter},'' \href{http://dx.doi.org/10.1103/PhysRevLett.121.161301}{{\em
  Phys. Rev. Lett.} {\bf 121} (2018) no.~16, 161301},
\href{http://arxiv.org/abs/1805.11753}{{\tt arXiv:1805.11753 [astro-ph.CO]}}.

\bibitem{Liu:2018icu}
H.~Liu, B.~D. Elwood, M.~Evans, and J.~Thaler, ``{Searching for Axion Dark
  Matter with Birefringent Cavities},''
  \href{http://dx.doi.org/10.1103/PhysRevD.100.023548}{{\em Phys. Rev. D} {\bf
  100} (2019) no.~2, 023548}, \href{http://arxiv.org/abs/1809.01656}{{\tt
  arXiv:1809.01656 [hep-ph]}}.

\bibitem{Berlin:2019ahk}
A.~Berlin, R.~T. D'Agnolo, S.~A.~R. Ellis, C.~Nantista, J.~Neilson,
  P.~Schuster, S.~Tantawi, N.~Toro, and K.~Zhou, ``{Axion Dark Matter Detection
  by Superconducting Resonant Frequency Conversion},''
\href{http://arxiv.org/abs/1912.11048}{{\tt arXiv:1912.11048 [hep-ph]}}.

\bibitem{Sikivie:2010fa}
P.~Sikivie, ``{Superconducting Radio Frequency Cavities as Axion Dark Matter
  Detectors},''
\href{http://arxiv.org/abs/1009.0762}{{\tt arXiv:1009.0762 [hep-ph]}}.

\bibitem{Bogorad:2019pbu}
Z.~Bogorad, A.~Hook, Y.~Kahn, and Y.~Soreq, ``{Probing Axionlike Particles and
  the Axiverse with Superconducting Radio-Frequency Cavities},''
  \href{http://dx.doi.org/10.1103/PhysRevLett.123.021801}{{\em Phys. Rev.
  Lett.} {\bf 123} (2019) no.~2, 021801},
  \href{http://arxiv.org/abs/1902.01418}{{\tt arXiv:1902.01418 [hep-ph]}}.

\bibitem{Lasenby:2019prg}
R.~Lasenby, ``{Microwave cavity searches for low-frequency axion dark
  matter},''
\href{http://arxiv.org/abs/1912.11056}{{\tt arXiv:1912.11056 [hep-ph]}}.

\bibitem{Romanenko:2014yaa}
A.~Romanenko, A.~Grassellino, A.~C. Crawford, D.~A. Sergatskov, and
  O.~Melnychuk, ``{Ultra-high quality factors in superconducting niobium
  cavities in ambient magnetic fields up to 190 mG},''
  \href{http://dx.doi.org/10.1063/1.4903808}{{\em Appl. Phys. Lett.} {\bf 105}
  (2014)  234103},
\href{http://arxiv.org/abs/1410.7877}{{\tt arXiv:1410.7877 [physics.acc-ph]}}.

\bibitem{Posen:2018bjn}
S.~Posen, G.~Wu, E.~Harms, A.~Grassellino, O.~S. Melnychuk, D.~A. Sergatskov,
  N.~Solyak, A.~Palczewski, D.~Gonnella, and T.~Peterson, ``{Role of magnetic
  flux expulsion to reach $Q_0 > 3 \times 10^{10}$ in superconducting rf
  cryomodules},''
  \href{http://dx.doi.org/10.1103/PhysRevAccelBeams.22.032001}{{\em Phys. Rev.
  Accel. Beams} {\bf 22} (2019) no.~3, 032001},
  \href{http://arxiv.org/abs/1812.03950}{{\tt arXiv:1812.03950
  [physics.acc-ph]}}.
[Phys. Rev. Accel. Beams22,032001(2019)].

\bibitem{DarkSRF}
A.~Grassellino, ``{SRF-based dark matter search: Experiment},''
\newblock 2020.
\newblock
  \url{https://indico.physics.lbl.gov/event/939/contributions/4371/attachments/2162/2915/DarkSRF-Aspen-2.pdf}.

\bibitem{Chaudhuri:2018rqn}
S.~Chaudhuri, K.~Irwin, P.~W. Graham, and J.~Mardon, ``{Fundamental Limits of
  Electromagnetic Axion and Hidden-Photon Dark Matter Searches: Part I - The
  Quantum Limit},''
\href{http://arxiv.org/abs/1803.01627}{{\tt arXiv:1803.01627 [hep-ph]}}.

\bibitem{datasheet}
Berkeley Nucleonics Corporation, {\em Model 865-M Wideband Synthesizer Data
  Sheet}, 5, 2019.
\newblock
  \url{https://www.berkeleynucleonics.com/sites/default/files/products/datasheets/865-m_datasheet_5-3-19_v1.04.pdf}.

\bibitem{Dicke:1946glx}
R.~H. Dicke, ``{The Measurement of Thermal Radiation at Microwave
  Frequencies},''
\href{http://dx.doi.org/10.1063/1.1770483}{{\em Rev. Sci. Instrum.} {\bf 17}
  (1946) no.~7, 268--275}.

\bibitem{Bernard:2001kp}
P.~Bernard, G.~Gemme, R.~Parodi, and E.~Picasso, ``{A Detector of small
  harmonic displacements based on two coupled microwave cavities},''
  \href{http://dx.doi.org/10.1063/1.1366636}{{\em Rev. Sci. Instrum.} {\bf 72}
  (2001)  2428--2437},
\href{http://arxiv.org/abs/gr-qc/0103006}{{\tt arXiv:gr-qc/0103006 [gr-qc]}}.

\bibitem{Bernard:2002ci}
P.~Bernard, A.~Chincarini, G.~Gemme, R.~Parodi, and E.~Picasso, ``{A Detector
  of gravitational waves based on coupled microwave cavities},''
\href{http://arxiv.org/abs/gr-qc/0203024}{{\tt arXiv:gr-qc/0203024 [gr-qc]}}.

\bibitem{Ballantini:2005am}
R.~Ballantini {\em et al.}, ``{Microwave apparatus for gravitational waves
  observation},''
\href{http://arxiv.org/abs/gr-qc/0502054}{{\tt arXiv:gr-qc/0502054 [gr-qc]}}.

\bibitem{Bernard:2000pz}
P.~Bernard, G.~Gemme, R.~Parodi, and E.~Picasso, ``{The RF control and
  detection system for PACO the parametric converter detector},''
  \href{http://arxiv.org/abs/physics/0004031}{{\tt arXiv:physics/0004031}}.

\bibitem{Foster:2017hbq}
J.~W. Foster, N.~L. Rodd, and B.~R. Safdi, ``{Revealing the Dark Matter Halo
  with Axion Direct Detection},''
  \href{http://dx.doi.org/10.1103/PhysRevD.97.123006}{{\em Phys. Rev. D} {\bf
  97} (2018) no.~12, 123006}, \href{http://arxiv.org/abs/1711.10489}{{\tt
  arXiv:1711.10489 [astro-ph.CO]}}.

\bibitem{Centers:2019dyn}
G.~P. Centers {\em et al.}, ``{Stochastic fluctuations of bosonic dark
  matter},''
\href{http://arxiv.org/abs/1905.13650}{{\tt arXiv:1905.13650 [astro-ph.CO]}}.

\bibitem{liu2017josephson}
G.~Liu, T.-C. Chien, X.~Cao, O.~Lanes, E.~Alpern, D.~Pekker, and M.~Hatridge,
  ``Josephson parametric converter saturation and higher order effects,'' {\em
  Applied Physics Letters} {\bf 111} (2017) no.~20, 202603.

\bibitem{Eriksson:2004cz}
D.~Eriksson, G.~Brodin, M.~Marklund, and L.~Stenflo, ``{A Possibility to
  measure elastic photon-photon scattering in vacuum},''
  \href{http://dx.doi.org/10.1103/PhysRevA.70.013808}{{\em Phys. Rev.} {\bf
  A70} (2004)  013808},
\href{http://arxiv.org/abs/physics/0411054}{{\tt arXiv:physics/0411054
  [physics]}}.

\bibitem{padamsee2009rf}
H.~Padamsee, {\em RF Superconductivity: Science, Technology, and Applications}.
\newblock John Wiley \& Sons, 2009.

\bibitem{rubiola2009phase}
E.~Rubiola, {\em Phase noise and frequency stability in oscillators}.
\newblock Cambridge University Press, 2009.

\bibitem{fnalex}
A.~Grassellino, ``{SRF-based dark matter search: Experiment},''
\newblock 2019.
\newblock
  \url{https://indico.fnal.gov/event/19433/session/2/contribution/2/material/slides/0.pdf}.

\bibitem{Saulson:2017jlf}
P.~R. Saulson, \href{http://dx.doi.org/10.1142/10116}{{\em {Fundamentals of
  Interferometric Gravitational Wave Detectors}}}.
\newblock World Scientific, 2nd. ed.~ed., 2017.

\bibitem{Meidlinger}
D.~Meidlinger, ``{A General Perturbation Theory for Cavity Mode Field
  Patterns},''
\newblock 2009.
\newblock \url{https://accelconf.web.cern.ch/SRF2009/papers/thppo005.pdf}.

\bibitem{Armengaud:2017nkf}
E.~Armengaud, N.~Palanque-Delabrouille, C.~Y{\`e}che, D.~J. Marsh, and J.~Baur,
  ``{Constraining the mass of light bosonic dark matter using SDSS
  Lyman-$\alpha$ forest},'' \href{http://dx.doi.org/10.1093/mnras/stx1870}{{\em
  Mon. Not. Roy. Astron. Soc.} {\bf 471} (2017) no.~4, 4606--4614},
  \href{http://arxiv.org/abs/1703.09126}{{\tt arXiv:1703.09126 [astro-ph.CO]}}.

\bibitem{Irsic:2017yje}
V.~Ir\v~si\v c, M.~Viel, M.~G. Haehnelt, J.~S. Bolton, and G.~D. Becker,
  ``{First constraints on fuzzy dark matter from Lyman-$\alpha$ forest data and
  hydrodynamical simulations},''
  \href{http://dx.doi.org/10.1103/PhysRevLett.119.031302}{{\em Phys. Rev.
  Lett.} {\bf 119} (2017) no.~3, 031302},
  \href{http://arxiv.org/abs/1703.04683}{{\tt arXiv:1703.04683 [astro-ph.CO]}}.

\bibitem{Nori:2018pka}
M.~Nori, R.~Murgia, V.~Ir\v~si\v c, M.~Baldi, and M.~Viel, ``{Lyman $\alpha$
  forest and non-linear structure characterization in Fuzzy Dark Matter
  cosmologies},'' \href{http://dx.doi.org/10.1093/mnras/sty2888}{{\em Mon. Not.
  Roy. Astron. Soc.} {\bf 482} (2019) no.~3, 3227--3243},
  \href{http://arxiv.org/abs/1809.09619}{{\tt arXiv:1809.09619 [astro-ph.CO]}}.

\bibitem{Garzilli:2019qki}
A.~Garzilli, O.~Ruchayskiy, A.~Magalich, and A.~Boyarsky, ``{How warm is too
  warm? Towards robust Lyman-$\alpha$ forest bounds on warm dark matter},''
  \href{http://arxiv.org/abs/1912.09397}{{\tt arXiv:1912.09397 [astro-ph.CO]}}.

\bibitem{Marsh:2018zyw}
D.~J. Marsh and J.~C. Niemeyer, ``{Strong Constraints on Fuzzy Dark Matter from
  Ultrafaint Dwarf Galaxy Eridanus II},''
  \href{http://dx.doi.org/10.1103/PhysRevLett.123.051103}{{\em Phys. Rev.
  Lett.} {\bf 123} (2019) no.~5, 051103},
  \href{http://arxiv.org/abs/1810.08543}{{\tt arXiv:1810.08543 [astro-ph.CO]}}.

\bibitem{Schutz:2020jox}
K.~Schutz, ``{Subhalo mass function and ultralight bosonic dark matter},''
  \href{http://dx.doi.org/10.1103/PhysRevD.101.123026}{{\em Phys. Rev. D} {\bf
  101} (2020) no.~12, 123026}, \href{http://arxiv.org/abs/2001.05503}{{\tt
  arXiv:2001.05503 [astro-ph.CO]}}.

\bibitem{graham2018spin}
P.~W. Graham, D.~E. Kaplan, J.~Mardon, S.~Rajendran, W.~A. Terrano, L.~Trahms,
  and T.~Wilkason, ``Spin precession experiments for light axionic dark
  matter,'' {\em Physical Review D} {\bf 97} (2018) no.~5, 055006.

\bibitem{wu2019search}
T.~Wu, J.~W. Blanchard, G.~P. Centers, N.~L. Figueroa, A.~Garcon, P.~W. Graham,
  D.~F.~J. Kimball, S.~Rajendran, Y.~V. Stadnik, A.~O. Sushkov, {\em et al.},
  ``Search for axionlike dark matter with a liquid-state nuclear spin
  comagnetometer,'' {\em Physical review letters} {\bf 122} (2019) no.~19,
  191302.

\bibitem{terrano2019constraints}
W.~A. Terrano, E.~G. Adelberger, C.~A. Hagedorn, and B.~R. Heckel,
  ``Constraints on axionlike dark matter with masses down to 10- 23 ev/c 2,''
  {\em Physical review letters} {\bf 122} (2019) no.~23, 231301.

\bibitem{abel2017search}
C.~Abel, N.~J. Ayres, G.~Ban, G.~Bison, K.~Bodek, V.~Bondar, M.~Daum,
  M.~Fairbairn, V.~V. Flambaum, P.~Geltenbort, {\em et al.}, ``Search for
  axionlike dark matter through nuclear spin precession in electric and
  magnetic fields,'' {\em Physical Review X} {\bf 7} (2017) no.~4, 041034.

\bibitem{Halverson:2019cmy}
J.~Halverson, C.~Long, B.~Nelson, and G.~Salinas, ``{Towards string theory
  expectations for photon couplings to axionlike particles},''
  \href{http://dx.doi.org/10.1103/PhysRevD.100.106010}{{\em Phys. Rev. D} {\bf
  100} (2019) no.~10, 106010}, \href{http://arxiv.org/abs/1909.05257}{{\tt
  arXiv:1909.05257 [hep-th]}}.

\bibitem{PetersonLNM}
U.~S.~G. Survey and J.~R. Peterson,
  \href{http://dx.doi.org/10.3133/ofr93322}{``Observations and modeling of
  seismic background noise,''} tech. rep., 1993.
\newblock \url{http://pubs.er.usgs.gov/publication/ofr93322}.

\bibitem{Gravimetry2018}
S.~Rosat and J.~Hinderer, ``Limits of detection of gravimetric signals on
  earth,'' \href{http://dx.doi.org/10.1038/s41598-018-33717-z}{{\em Scientific
  Reports} {\bf 8} (2018) no.~1, 15324}.
  \url{https://doi.org/10.1038/s41598-018-33717-z}.

\bibitem{Wolf:2002ip}
P.~Wolf, S.~Bize, A.~Clairon, A.~N. Luiten, G.~Santarelli, and M.~E. Tobar,
  ``{Tests of relativity using a microwave resonator},''
  \href{http://dx.doi.org/10.1103/PhysRevLett.90.060402}{{\em Phys. Rev. Lett.}
  {\bf 90} (2003)  060402}, \href{http://arxiv.org/abs/gr-qc/0210049}{{\tt
  arXiv:gr-qc/0210049}}.

\bibitem{Stanwix:2005yv}
P.~L. Stanwix, M.~E. Tobar, P.~Wolf, M.~Susli, C.~R. Locke, E.~N. Ivanov,
  J.~Winterflood, and F.~van Kann, ``{Test of Lorentz invariance in
  electrodynamics using rotating cryogenic sapphire microwave oscillators},''
  \href{http://dx.doi.org/10.1103/PhysRevLett.95.040404}{{\em Phys. Rev. Lett.}
  {\bf 95} (2005)  040404}, \href{http://arxiv.org/abs/hep-ph/0506074}{{\tt
  arXiv:hep-ph/0506074}}.

\bibitem{Stanwix:2006jb}
P.~L. Stanwix, M.~E. Tobar, P.~Wolf, C.~R. Locke, and E.~N. Ivanov, ``{Improved
  test of Lorentz invariance in electrodynamics using rotating cryogenic
  sapphire oscillators},''
  \href{http://dx.doi.org/10.1103/PhysRevD.74.081101}{{\em Phys. Rev. D} {\bf
  74} (2006)  081101}, \href{http://arxiv.org/abs/gr-qc/0609072}{{\tt
  arXiv:gr-qc/0609072}}.

\bibitem{RevModPhys.82.1155}
A.~A. Clerk, M.~H. Devoret, S.~M. Girvin, F.~Marquardt, and R.~J. Schoelkopf,
  \href{http://dx.doi.org/10.1103/RevModPhys.82.1155}{``Introduction to quantum
  noise, measurement, and amplification,''{\em Rev. Mod. Phys.} {\bf 82} (Apr,
  2010)  1155--1208}.
  \url{https://link.aps.org/doi/10.1103/RevModPhys.82.1155}.

\bibitem{Arvanitaki:2019rax}
A.~Arvanitaki, S.~Dimopoulos, M.~Galanis, L.~Lehner, J.~O. Thompson, and
  K.~Van~Tilburg, ``{Large-misalignment mechanism for the formation of compact
  axion structures: Signatures from the QCD axion to fuzzy dark matter},''
  \href{http://dx.doi.org/10.1103/PhysRevD.101.083014}{{\em Phys. Rev. D} {\bf
  101} (2020) no.~8, 083014}, \href{http://arxiv.org/abs/1909.11665}{{\tt
  arXiv:1909.11665 [astro-ph.CO]}}.

\bibitem{Fairbairn:2017sil}
M.~Fairbairn, D.~J.~E. Marsh, J.~Quevillon, and S.~Rozier, ``{Structure
  formation and microlensing with axion miniclusters},''
  \href{http://dx.doi.org/10.1103/PhysRevD.97.083502}{{\em Phys. Rev. D} {\bf
  97} (2018) no.~8, 083502}, \href{http://arxiv.org/abs/1707.03310}{{\tt
  arXiv:1707.03310 [astro-ph.CO]}}.

\bibitem{Kolb:1993zz}
E.~W. Kolb and I.~I. Tkachev, ``{Axion miniclusters and Bose stars},''
  \href{http://dx.doi.org/10.1103/PhysRevLett.71.3051}{{\em Phys. Rev. Lett.}
  {\bf 71} (1993)  3051--3054}, \href{http://arxiv.org/abs/hep-ph/9303313}{{\tt
  arXiv:hep-ph/9303313}}.

\bibitem{Kolb:1993hw}
E.~W. Kolb and I.~I. Tkachev, ``{Nonlinear axion dynamics and formation of
  cosmological pseudosolitons},''
  \href{http://dx.doi.org/10.1103/PhysRevD.49.5040}{{\em Phys. Rev. D} {\bf 49}
  (1994)  5040--5051}, \href{http://arxiv.org/abs/astro-ph/9311037}{{\tt
  arXiv:astro-ph/9311037}}.

\bibitem{Kolb:1994fi}
E.~W. Kolb and I.~I. Tkachev, ``{Large amplitude isothermal fluctuations and
  high density dark matter clumps},''
  \href{http://dx.doi.org/10.1103/PhysRevD.50.769}{{\em Phys. Rev. D} {\bf 50}
  (1994)  769--773}, \href{http://arxiv.org/abs/astro-ph/9403011}{{\tt
  arXiv:astro-ph/9403011}}.

\bibitem{Kolb:1995bu}
E.~W. Kolb and I.~I. Tkachev, ``{Femtolensing and picolensing by axion
  miniclusters},'' \href{http://dx.doi.org/10.1086/309962}{{\em Astrophys. J.
  Lett.} {\bf 460} (1996)  L25--L28},
  \href{http://arxiv.org/abs/astro-ph/9510043}{{\tt arXiv:astro-ph/9510043}}.

\bibitem{Cowan:2010js}
G.~Cowan, K.~Cranmer, E.~Gross, and O.~Vitells, ``{Asymptotic formulae for
  likelihood-based tests of new physics},''
  \href{http://dx.doi.org/10.1140/epjc/s10052-011-1554-0}{{\em Eur. Phys. J. C}
  {\bf 71} (2011)  1554}, \href{http://arxiv.org/abs/1007.1727}{{\tt
  arXiv:1007.1727 [physics.data-an]}}. [Erratum: Eur.Phys.J.C 73, 2501 (2013)].

\bibitem{Wilks:1938dza}
S.~Wilks, ``{The Large-Sample Distribution of the Likelihood Ratio for Testing
  Composite Hypotheses},''
  \href{http://dx.doi.org/10.1214/aoms/1177732360}{{\em Annals Math. Statist.}
  {\bf 9} (1938) no.~1, 60--62}.

\bibitem{Chaudhuri:2014dla}
S.~Chaudhuri, P.~W. Graham, K.~Irwin, J.~Mardon, S.~Rajendran, and Y.~Zhao,
  ``{Radio for hidden-photon dark matter detection},''
  \href{http://dx.doi.org/10.1103/PhysRevD.92.075012}{{\em Phys. Rev.} {\bf
  D92} (2015) no.~7, 075012},
\href{http://arxiv.org/abs/1411.7382}{{\tt arXiv:1411.7382 [hep-ph]}}.

\end{thebibliography}\endgroup

\clearpage
\newpage
\maketitle
\onecolumngrid
\begin{center}
\textbf{\large Heterodyne Broadband Detection of Axion Dark Matter} \\ 
\vspace{0.05in}
{ \it \large Supplemental Material}\\ 
\vspace{0.05in}
{}
{Asher Berlin, Raffaele Tito D'Agnolo, Sebastian A. R. Ellis, and Kevin Zhou}

\end{center}
\setcounter{equation}{0}
\setcounter{figure}{0}
\setcounter{table}{0}
\setcounter{section}{1}
\renewcommand{\theequation}{S\arabic{equation}}
\renewcommand{\thefigure}{S\arabic{figure}}
\renewcommand{\thetable}{S\arabic{table}}
\interfootnotelinepenalty=10000 

\setstretch{1.1}

In this Supplemental Material, we derive the experimental sensitivity of our proposed approach in detail. We begin by deriving the signal and noise PSDs. We then discuss the statistical procedure used to estimate the reach, which involves subtleties for integration times shorter than the axion coherence time. The last page of the Supplemental Material contains a table which summarizes our notation and can be of use to a reader who wants to carefully follow the derivations below. 


\section*{Definitions and Conventions}

Throughout, we use the conventions of Ref.~\cite{Berlin:2019ahk}. In particular, the Fourier transform of a function $f (t)$ is denoted as $f(\w)$, where
\be
f(t) = \frac{1}{2 \pi} \, \int d \w ~ e^{i \w t} f( \w) 
~,~ 
f(\w) = \int dt ~ e^{-i \w t} f(t)
\, .
\ee
The two-sided PSD of $f$, denoted as $S_f(\w)$, is defined to be
\be
\label{eq:PSDdef}
\langle f(\w) f^*(\w^\p) \rangle = S_f(\w) \, \delta (\w - \w^\p)
~,
\ee
where the brackets correspond to an ensemble average. The steady state time-averaged power is then given by
\be
\label{eq:AvgPwr}
\langle \, f (t)^2 \, \rangle = \frac{1}{ (2 \pi)^2} \, \int d \w ~ S_f(\w)
~,
\ee
where all integrals over $t$ or $\w$ are taken from $-\infty$ to $\infty$, unless specified otherwise. 

The fields in the pump ($i = 0$) and signal ($i = 1$) modes behave as damped driven harmonic oscillators. Therefore, we will find it convenient to treat them as independent RLC circuits of resonant frequency $\w_i = 1 / \sqrt{L_i C_i}$ and quality factor $Q_i = \sqrt{L_i / C_i} \, / R_i \propto 1/R_i$. This is not a physical statement, but merely a mathematical analogy between two systems obeying the same equations. Furthermore, we often approximate $\w_1 \simeq \w_0$, unless the difference $\w_1 - \w_0$ is important, in which case we leave the expression generalized to $\w_1 \neq \w_0$. 

It is also useful to recall the distinction between the intrinsic and loaded quality factors of a cavity mode. The quality factors of the pump and signal modes are denoted by $Q_0$ and $Q_1$, respectively, and $Q_1$ is determined by both the intrinsic energy loss of the cavity $Q_{\text{int}} \simeq Q_0 \gtrsim 10^{10}$ and the coupling to the readout $Q_\text{cpl}$, 
\be 
\label{eq:Qtot}
\frac{1}{Q_1} = \frac{1}{Q_\text{int}} + \frac{1}{Q_\text{cpl}}
~.
\ee
In the RLC circuit analogy, this corresponds to the addition of resistances in series, $R_1 = R_{\text{int}} + R_{\text{cpl}}$. As discussed in the main body, it can be beneficial to overcouple, such that $Q_1 \simeq Q_\text{cpl} \ll Q_\text{int}$. For simplicity, we begin by deriving the noise and signal PSDs corresponding to the \emph{total} power delivered to the cavity. However, the sensitivity of the apparatus depends only on the power delivered to the readout, and when discussing overcoupling, we will explicitly show how the PSDs must be modified to account for this.

The total noise PSD $S_n$ receives contributions from leakage noise, mechanical mode mixing, thermal noise, and amplifier noise,
\be
S_n(\w)=S_\text{leak} (\w)+S_{\rm mix}(\w)+S_{\rm th}(\w)+S_{\rm amp}(\w)\, .
\ee
Leakage noise from the oscillator is the dominant noise source at low axion masses and is parametrized as
\be
\label{eq:Sdrive}
S_\text{leak} (\w) =  \eps^2 \, P_\text{in} \, \left( S_{b_0} (\w) + \frac{Q_1}{Q_0} \, S_{b_1} (\w) \right) \equiv S_\text{leak}^{(0)} (\w) + S_\text{leak}^{(1)} (\w)
~,
\ee
where
\be
\label{eq:Pin}
P_\text{in} \equiv (\w_0 / Q_0) \, B_0^2 \, V_\text{cav}
\ee
is the power stored in the cavity and the $b_i$ are defined below. Both terms in \Eq{Sdrive} are suppressed by $\epsilon \ll 1$, which parametrizes the cross-coupling between the pump mode and readout waveguide, and the coupling between the signal mode and the loading waveguide, which are of the same order.

We define the characteristic amplitude of the pump and signal mode magnetic fields as
\be
\label{eq:B0}
B_i \equiv \sqrt{\frac{1}{V_\text{cav}} \int_{V_\text{cav}} |\Bvec_i (x)|^2} \label{eq:Bfield}
~,
\ee
where $\Bvec_i (x)$ is the time-independent part of the magnetic field
\be
\Bvec_i (x,t)=\Bvec_i (x) \, b_i(t)
\, ,
\ee
and $b_i (t)$ is the dimensionless time-dependent coefficient. For instance, for a monochromatic source exciting mode $i$, $b_i(t) = \cos{\w_i t}$ and $S_{b_i} (\w) = \pi^2 \, \Big( \delta(\w - \w_i) + \delta(\w+\w_i) \Big)$. More generally, phase noise of the oscillator and mechanical vibrations contribute to $S_{b_i} (\w)$ such that
\be
\label{eq:Sbidef}
S_{b_i} (\w) = \pi^2 \, \Big( \delta(\w - \w_i) + \delta(\w+\w_i) \Big) + S_{b_i}^{(\text{phase})} + S_{b_i}^{(\text{mech})}
~,
\ee
where $ S_{b_i}^{(\text{phase})}$ and $ S_{b_i}^{(\text{mech})}$ are given by Eqs.~(\ref{eq:Sbi}) and (\ref{eq:SbiMech2}), respectively. 

\section*{Leakage Noise and Signal}

\subsection{Oscillator Phase Noise}

We model the oscillator as a voltage $V_d$ that drives the equivalent RLC circuits of the pump and signal modes. In particular, a noisy oscillator is parametrized as a driving voltage with a time-dependent phase $\varphi(t)$,
\be
V_d (t) = V_d \, \cos{(\w_0 t + \ph(t))}
~,
\ee
where the voltage amplitude is fixed to the power delivered to the pump mode,
\be
\label{eq:V0}
V_d^2 / R_0 = P_\text{in}
~.
\ee
When the amplitude of the phase is small ($\ph(t) \ll 1$), the above form can be expanded as
\be
V_d (t) \simeq V_d  \, \big( \cos{\w_0 t}  - \ph(t) \sin{\w_0 t} \, \big)
~.
\ee
This implies that the PSD of the drive voltage is
\be
S_{V_d} (\w) \simeq R_0 \,  P_\text{in} \, \Big[ \pi^2 \, \Big( \delta(\w - \w_0) + \delta(\w+\w_0) \Big) + \frac{1}{4} \, \Big( S_\ph (\w - \w_0) + S_\ph (\w + \w_0) \Big) \Big]
~.
\ee
As we discuss below, $S_\ph (\w)$ is peaked near $\w \simeq 0$ (see \Eq{cphase}). Therefore, for a frequency $\w$ of fixed sign, only one of $S_\ph (\w \mp \w_0)$ dominates in the expression above. By convention, we focus on $\w \simeq \w_0 > 0$, such that $S_\ph (\w - \w_0) \gg S_\ph (\w + \w_0)$. Also using that $S_\ph (\w) = S_\ph (-\w) = S_\ph (|\w|)$, we then have
\be
\label{eq:SVd}
S_{V_d} (\w) \simeq R_0 \,  P_\text{in} \, \Big[ \pi^2 \, \Big( \delta(\w - \w_0) + \delta(\w+\w_0) \Big) + \frac{1}{4} \, S_\ph (|\w - \w_0 |) \Big]
~.
\ee
Above, the first two terms involving delta functions are simply the PSD of a perfectly monochromatic drive. The inclusion of $S_\ph$ accounts for so-called ``phase noise" of an imperfect oscillator. 

The power delivered to the $i^\text{th}$ mode is determined by the voltage $V_i$ across the resistor $R_i$, which obeys Kirchoff's voltage law,
\be
\ddot V_i(t) +\frac{\w_i}{Q_i} \, \dot V_i(t)+\w_i^2 \, V_i(t)=\frac{\w_i}{Q_i} \, \dot V_d(t) 
\, .
\ee
In the above equation, we have not included the fact that the oscillator's coupling to the $i = 1$ signal mode is suppressed by $\eps$. For convenience, we have instead included this factor in \Eq{Sdrive}, so that the derivation of $S_\text{leak}^{(i)}$ is identical for $i = 0$ and $i=1$. Fourier transforming and solving for the PSD of $V_i$ gives the cavity response function,
\be
\label{eq:SVR}
S_{V_i} (\w) = \frac{(\w \, \w_i / Q_i)^2}{(\w^2 - \w_i^2)^2 + (\w \, \w_i / Q_i)^2} ~ S_{V_d} (\w)
~.
\ee
To change variables from $V_i$ to $b_i$, we equate the total power, $S_{V_i} (\w) / R_i = (Q_i / Q_0) \,  P_\text{in} \,  S_{b_i} (\w)$, giving
\be
\label{eq:Sbi}
\eqbox{
S_{b_i}^{(\text{phase})} (\w) \simeq \frac{1}{4} ~ \frac{(\w \, \w_0 / Q_i)^2}{(\w^2 - \w_i^2)^2 + (\w \, \w_0 / Q_i)^2} ~ S_\ph (|\w - \w_0 |)
}
~,
\ee
where we approximated $\w_0 / Q_i \simeq \w_1 / Q_i$.

To incorporate the small, but finite, width of the external oscillator ($\Delta \w_d \lesssim \text{mHz}$), we replace the delta functions in \Eq{Sbidef} by
\be
\label{eq:DeltaDriver}
\delta (\w) \simeq \frac{\Theta(\Delta \w_d / 2 - |\w|)}{\Delta \w_d}
~,
\ee
where $\Theta$ is the Heaviside step function. For simplicity, we make the approximation $\Delta \w_d \ll \w_0 / Q_\text{int}$ throughout our calculations, since this holds for all parameters we consider. As in Ref.~\cite{Berlin:2019ahk}, for the phase noise PSD $S_\ph (\w)$, we fit the reported spectrum of a commercially available oscillator~\cite{datasheet} to the form 
\be
\label{eq:cphase}
S_\ph(\w) = \sum_{n=0}^3 c_n \, \w^{-n}
\, .
\ee
We find that the coefficients
\be
c_0 \sim 10^{-15}\ \Hz^{-1}
~,~
c_1 \sim 10^{-11}
~,~
c_2 \sim 10^{-10}\ \Hz
~,~
c_3 \sim 10^{-8}\ \Hz^2
~,
\ee
provide a good fit for $\w_0 \sim 100 \ \MHz$. We fix the overall normalization by demanding that the phase noise term of \Eq{Sbi} smoothly matches on to the the central peak of $S_{b_i} (\w)$ near $\w \simeq \w_0$ when using \Eq{DeltaDriver} and $\Delta \w_d \simeq 0.1 \ \mHz$. 

\subsection{Mechanical Leakage Noise}

An additional contribution to $S_{b_i}$ arises from small mechanical vibrations of the cavity walls, which lead to time-dependent shifts of the resonant frequencies. These frequency wobbles affect the mode PSDs by enhancing the power in the high frequency tail. We incorporate this effect by continuing with the analogy to an RLC circuit. If the resonant frequency squared of an equivalent RLC circuit has a small fractional time variation $\Delta (t)$, Kirchoff's voltage law becomes
\be
\label{eq:KirchoffMod}
\ddot V_i(t) +\frac{\w_i}{Q_i} \, \dot V_i(t)+\w_i^2 V_i(t)=\frac{\w_i}{Q_i} \, \dot V_d(t) - \omega_i^2 \, \Delta(t) \, V_i(t) 
\, .
\ee
Solving this equation to first order in $\Delta$ yields
\be
\label{eq:SVR2}
S_{V_i} (\w) \simeq \frac{(\w \, \w_i / Q_i)^2}{(\w^2 - \w_i^2)^2 + (\w \, \w_i / Q_i)^2} \left( S_{V_d} (\w) + \frac{\w_i^4}{(2 \pi)^2} \int d \w^\p ~ \frac{S_\Delta (\w - \w^\p) \, S_{V_d} (\w^\p)}{(\w^{\p \, 2} - \w_i^2)^2 + (\w^\p \, \w_i / Q_i)^2} \right)
~.
\ee
Relative to \Eq{SVR}, the second term incorporates perturbative corrections from mechanical vibrations. Substituting the leading order piece of $S_{V_d} (\w)$ from \Eq{SVd} into the $\w^\p$ integral of \Eq{SVR2} and converting from $V_i$ to $b_i$ again yields the mechanical vibration contribution to the unit-normalized mode PSD, 
\be
\label{eq:SbiMech1}
S_{b_i}^{(\text{mech})} (\w) \simeq \frac{1}{4} \, \frac{(\w \, \w_0 / Q_i)^2}{(\w^2 - \w_i^2)^2 + (\w \, \w_0 / Q_i)^2} ~ 
\frac{\w_0^4}{(\w_0^2 - \w_i^2)^2 + (\w_0^2 / Q_i)^2} ~ \big( S_\Delta (\w - \w_0) + S_\Delta (\w + \w_0) \big) ~.
\ee
When the modes are degenerate, the second Breit--Wigner factor above simplifies, giving
\be
\label{eq:SbiMech1Simple}
S_{b_i}^{(\text{mech})} (\w) \simeq \frac{1}{4} \, 
\frac{(\w \, \w_0)^2}{(\w^2 - \w_i^2)^2 + (\w \, \w_0 / Q_i)^2} ~ \big( S_\Delta (\w - \w_0) + S_\Delta (\w + \w_0) \big) ~.%
\ee
For brevity, we will use this form below, though the more general form \Eq{SbiMech1} is useful when considering what happens when the modes are not exactly degenerate.

The PSD of the frequency wobble $S_\Delta (\w)$ can be computed using cavity perturbation theory, which treats the small displacement of the cavity walls as an expansion parameter. We will assume that for each axion mass, a single mechanical resonance, labeled by ``$m$," dominates the mechanical vibrations. To first order in cavity perturbation theory, $\Delta (t) \simeq - q_m (t) \, C_i^m$, where the displacement of the cavity walls, projected onto the spatial profile of the mechanical resonance, is parametrized by the generalized coordinate $q_m$, and the coupling coefficient $C_i^m$ quantifies the mechanical overlap of the electromagnetic cavity modes ($i = 0,1$) with the $m^\text{th}$ vibrational mode of the cavity walls~\cite{Bernard:2002ci,Berlin:2019ahk}. Parametrically, fractional length variations are 
comparable to the fractional frequency variations they induce, so $|C_i^m| \sim V_\text{cav}^{-1/3}$ for maximally coupled mechanical and electromagnetic modes. 

The amplitude of the wall displacement $q_m$ is determined by the generalized force $f_m$, such that
\be \label{eq:elastic_deformation_eq}
S_{q_m} (\w) = \frac{S_{f_m} (\w) / M_\text{cav}^2}{(\w^2 - \w_m^2)^2 + (\w \, \w_m / Q_m)^2}
~,
\ee
where $M_\text{cav}$ is the mass of the cavity, $\w_m$ is the frequency of the mechanical resonance, and $Q_m$ is its corresponding mechanical quality factor. Here, $f_m$ should be regarded as the remaining force that couples to the cavity after vibrational attenuation is employed. Since $S_\Delta = |C_i^m|^2 \, S_{q_m}$, 
\be
\label{eq:SDelta1}
S_\Delta (\w \pm \w_0) = \frac{|C_i^m|^2 \, S_{f_m} (\w \pm \w_0) / M_\text{cav}^2}{\big((\w \pm \w_0 \big)^2 - \w_m^2)^2 + \big((\w \pm \w_0) \, \w_m / Q_m \big)^2}
~.
\ee
The PSD of the generalized force $f_m$ is peaked towards smaller $\w$~\cite{Ballantini:2005am}, which implies that for frequencies near $\w \simeq \w_0$ the $S_\Delta (\w - \w_0)$ term dominates over the $S_\Delta (\w + \w_0)$ term in \Eq{SbiMech1Simple}. As in Ref.~\cite{Berlin:2019ahk}, we determine the size of the force PSD by fixing the RMS cavity wall displacement $q_\text{rms}\simeq 0.1 \ \nm$, consistent with DarkSRF~\cite{DarkSRF}, and assume that it is dominated by the lowest-lying mechanical resonance of the cavity, with corresponding frequency $\w_\text{min}$: 
\be
\label{eq:Sfm}
S_{f_m} (\w_\text{min}) \simeq 4 \pi \, M_\text{cav}^2 \, \w_\text{min}^3 \, q_\text{rms}^2 / Q_m \simeq 10^{-13} \ \text{N}^2 \ \Hz^{-1} \times \bigg( \frac{M_\text{cav}}{1\ \text{kg}} \bigg)^{2} \bigg( \frac{\w_\text{min}}{1\ \kHz} \bigg)^{3} \bigg( \frac{q_\text{rms}}{0.1 \, \nm} \bigg)^{2} \bigg( \frac{10^3}{Q_m}\bigg)
~.
\ee
In the following discussion, we take $\w_\text{min} = 1\ \kHz$ and $Q_m = 10^3$, which are representative of the SRF cavities fabricated for the MAGO experiment~\cite{Bernard:2001kp,Ballantini:2005am}. For maximally coupled mechanical and electromagnetic modes, Eqs.~(\ref{eq:SbiMech1Simple}), (\ref{eq:SDelta1}), and (\ref{eq:Sfm}) imply that 
\be
\label{eq:SbiMech2}
\eqbox{
S_{b_i}^{(\text{mech})} (\w) \simeq \frac{ (\w \, \w_0)^2}{(\w^2 - \w_i^2)^2 + (\w \, \w_0 / Q_i)^2} ~ \frac{ \pi \, \w_\text{min}^3 \, \delta^2 / Q_m}{((\w - \w_0)^2 - \w_m^2)^2 + ((\w - \w_0) \, \w_m / Q_m)^2} 
}
~,
\ee
where we defined the fractional cavity wall displacement $\delta \equiv q_{\text{rms}} / V_{\text{cav}}^{1/3}$. For our baseline estimates, we take $q_\text{rms} = 0.1 \ \nm$, corresponding to $\delta \sim 10^{-10}$ for a meter-sized cavity.

In the SRF cavity setup of Ref.~\cite{Bernard:2001kp}, direct measurements found a forest of mechanical resonances above $\w_\text{min}$, approximately separated by $100 \, \Hz$. For each axion mass $m_a$ above $\w_{\text{min}}$, mechanical noise is most severe if there exists a resonance at the axion mass, $\w_m \simeq m_a$, and is least severe if the nearest resonance is separated by $50 \ \Hz$. We thus estimate the median noise PSD for each value of $m_a$ by taking the nearest mechanical resonance to be separated by $25 \ \Hz$,
\be
\label{eq:wmech}
\w_m (m_a) \simeq \text{max}\left(\w_{\text{min}}, m_a + 25 \, \Hz \right)
.
\ee
See Sec.~VC of Ref.~\cite{Berlin:2019ahk} for a more detailed discussion regarding this point. 

To estimate $S_{f_m} (\w)$ at lower frequencies, we assume that the attenuated $S_{f_m} (\w)$ that enters our calculations is flat, i.e., $S_{f_m} (\w) \simeq S_{f_m} (\w_\text{min})$, though the precise spectral shape will depend on the details of the vibration attenuation mechanism. This estimate is consistent with measured \textit{unattenuated} acceleration PSDs from seismic activity at frequencies as low as $10\ \mu\text{Hz}$~\cite{PetersonLNM, Gravimetry2018}. Thus, given the implementation of even modest seismic isolation, our estimate for the low frequency force PSD is quite possibly pessimistic. 

The effects of vibrations at very low frequencies, $\w \lesssim \Delta \w_r$, are resonantly enhanced by the Breit--Wigner factor in \Eq{SbiMech1Simple}. This can cause our perturbative calculation to break down even though $\delta$ is small. To ensure this is not an issue, we demand that $S_{b_i}^{(\text{mech})}$ in \Eq{SbiMech2} is smaller than the leading order terms in \Eq{Sbidef}. Approximating the delta functions as in \Eq{DeltaDriver}, this condition holds at low frequencies if
\be
\label{eq:deltaperturbative}
\delta \lesssim \left( \frac{Q_m ~ \w_\text{min}}{Q_i^2 ~ \Delta \w_d} \right)^{1/2} \sim 10^{-7} \times \left( \frac{10^{12}}{Q_i} \right)
~,
\ee
which is easily satisfied. In fact, the sensitivity of our setup is even robust to $\delta$ near the perturbative limit, as shown in the lower-left panel of \Fig{reachvary}, corresponding to vibrational forces many orders of magnitude greater than seismic noise.

It is worth comparing this situation to that faced by interferometric experiments, where seismic noise is an important limiting factor at low frequencies. Such experiments precisely measure the distance between multiple objects, which are typically freely hung and are independently subject to seismic vibrations. By contrast, our approach takes place entirely within a single rigid cavity. Only the \textit{relative} motion between the cavity walls is relevant for noise, and this is many orders of magnitude smaller than the RMS motion of the ground itself. However, throughout this section, we have used \Eq{elastic_deformation_eq}, which assumes that the mechanical response of the cavity is elastic. Nonelastic deformations can lead to slow drifts of the cavity frequencies, which is addressed in a dedicated section below.

\subsection*{Signal Power}

We calculate the signal PSD using the drive mode PSD $S_{b_0} (\w)$ derived above. From Eq.~(19) of Ref.~\cite{Berlin:2019ahk}, the general form for the signal PSD is
\be
\label{eq:Ssig1}
S_\text{sig} (\w) = \frac{1}{(2 \pi)^2} \, \frac{\w_0}{Q_1} \, \left( \gyy \, \eta_a \, B_0 \right)^2 V_\text{cav} ~ \frac{\w^2 \int d \w^\p ~ I_a(\w, \w^\p)}{(\w^2 - \w_1^2)^2 + (\w \, \w_0 / Q_1)^2}
~,
\ee
where the axion form factor $\eta_a$ is defined as in Eq.~\eqref{eq:formfactor} and
\be
I_a (\w , \w^\p) \equiv (\w - \w^\p)^2 \, S_a (\w - \w^\p) \, S_{b_0} (\w^\p) 
~.
\ee
Above, $S_a (\w)$ is the PSD of the axion field, and $S_{b_0} (\w)$ includes contributions from Eqs.~(\ref{eq:Sbi}) and (\ref{eq:SbiMech2}), as in \Eq{Sbidef}. For the axion PSD, we use a simplified form that neglects effects from solar and terrestrial motion,
\be
\label{eq:Sa}
S_a (\w) = \Theta (|\w| - m_a) ~ \frac{2 \pi^2 \, \rhodm}{m_a^3 \, \sigma_v^2} \, e^{- (|\w| - m_a)/(m_a \, \sigma_v^2)}
~,
\ee
where the dispersion velocity is $\sigma_v \simeq 9 \times 10^{-4}$. This is consistent with the normalization $\langle a^2 \rangle = \rhodm / m_a^2$. Examples of the signal PSD, compared to the total noise PSD, are shown in \Fig{PSD} for various values of $m_a$. 

When there is a large hierarchy between the widths of the external oscillator and the axion field, \Eq{Ssig1} can be simplified by analytically evaluating the $\w^\p$ integral involving $I_a (\w, \w^\p)$. For instance, when the axion is much narrower, $\Delta \w_a \ll \Delta \w_d$, we use
\be
S_a(\w) \simeq (2 \pi)^2 \, \frac{\rhodm}{2 m_a^2} \, \Big( \delta (\w - m_a) + \delta(\w+m_a) \Big)
\ee
in \Eq{Ssig1}. Instead, if the oscillator is narrower, $\Delta \w_d \ll \Delta \w_a$, then we use
\be
S_{b_0} (\w) \simeq \pi^2 \, \Big( \delta(\w - \w_0) + \delta(\w+\w_0) \Big) ~.
\ee
In these limits, the signal PSD simplifies to 
\begin{empheq}[box=\fcolorbox{light-gray}{light-gray}]{align}
\label{eq:SsigApprox}
S_\text{sig} (\w) \simeq
\frac{1}{2} \, \frac{\w_0}{Q_1} \, \left( \gyy \, \eta_a \, B_0 \right)^2 \, V_\text{cav} \times
\begin{cases}
 \rhodm \, \frac{\w^2 \big( S_{b_0} (\w - m_a) + S_{b_0} (\w+m_a) \big)}{(\w^2 - \w_1^2)^2 + (\w \, \w_0 / Q_1)^2 } & \Delta \w_a \ll \Delta \w_d \, ,
\medskip
\\
\frac{1}{2} ~ \frac{\w^2 \big( (\w - \w_0)^2 \, S_a(\w - \w_0) + (\w + \w_0)^2 \, S_a(\w + \w_0) \big)}{(\w^2 - \w_1^2)^2 + (\w \, \w_0 / Q_1)^2 } & \Delta \w_d \ll \Delta \w_a
~.
\end{cases}
\end{empheq}
The expression for the signal power in \Eq{Psig1} can be obtained by approximating the narrowest piece of \Eq{SsigApprox} ($S_a (\w)$ of width $\Delta \w_a$, $S_{b_0} (\w)$ of width $\Delta \w_d$, or the cavity resonance of width $\Delta \w_r$) in the expression above as a delta function and integrating over $\w$. 

\section*{Additional Noise Sources}

\subsection*{Mechanical Noise from Mode Mixing}

In the previous section, we showed that mechanical vibrations contribute to leakage noise by affecting how the external oscillator loads power into the high frequency tail of the pump and signal mode PSDs. In addition, deformations of the cavity can lead to ``mode mixing," thus allowing for \textit{direct} power transfer between the two modes of interest. 

To describe this effect, we use the cavity perturbation theory results of Refs.~\cite{Meidlinger,Bernard:2002ci}. For a single mechanical resonance, labeled by ``$m$," to leading order in the fractional displacement of the cavity wall $\delta$, the equation of motion governing the time-evolution of the signal mode is
\be
\label{eq:modemix2}
\ddot V_1(t) + \frac{\w_1}{Q_1} \, \dot V_1(t) + \w_1^2 \, V_1(t) = \w_1^2 \, \eta_\text{mix} \, \delta(t) \, V_0(t)
~,
\ee
where we again have used the analogy to an RLC circuit. The dimensionless mechanical form factor $\eta_{\text{mix}}$ is
\be 
\label{eq:etamix}
\eta_\text{mix} \propto \int d \Sv \cdot \xiv_m (x) ~ \big( \Evec_0 (x) \cdot \Evec_1 (x) - \Bvec_0 (x) \cdot \Bvec_1 (x) \big)
~,
\ee
where the integral is performed over the surface of the cavity and the spatial profile of the mechanical mode is characterized by the normalized mode vector $\xiv_m$. For a perfectly cylindrical cavity, the pump and signal modes considered in this work are locally orthogonal, and so $\eta_{\text{mix}} = 0$. However, in reality the cavity cannot be manufactured perfectly, and its shape continues to change throughout the experiment due to low frequency deformations sourced by, e.g., seismic noise or fluctuations in the ambient temperature. 

We parametrize these static and slowly varying deviations from a cylindrical shape with a fractional displacement $\delta_s(t)$, which has support only on frequencies much less than $\omega_{\text{min}} \sim \kHz$. Now, $\eta_{\text{mix}}(t) \sim \delta_s(t)$ and the perturbative correction to $S_{V_1}$ from mode mixing in \Eq{modemix2} is precisely the same as that of leakage noise in \Eq{KirchoffMod}, except that the driving term is proportional to $\eta_{\text{mix}} \, \delta(t) \, V_0(t)$ rather than $C_1^m \, q_m(t) \, V_1(t) \sim \delta(t) \, V_1(t)$.\footnote{Alternatively, one could include the next order term on the RHS of \Eq{modemix2} as $\eta_\text{mix} \, \delta(t) \to \eta_\text{mix} \, \delta(t) + \tilde{\eta}_\text{mix} \, \delta^2(t)$ where $\tilde{\eta}_\text{mix}$ is an $\order{1}$ form factor that does not vanish even for locally orthogonal modes. Then, upon decomposing $\delta(t)$ in terms of slow and fast components as $\delta(t) = \delta_s (t) + \delta_f (t)$, including the cross term $\delta_s \delta_f$ is equivalent to simply including this contribution as $\eta_\text{mix} \sim \delta_s$, as we have done here.} Thus, by following the same logic as was used to derive \Eq{SbiMech2}, the noise PSD $S_\text{mix}$ from mode mixing is parametrically
\be
\frac{S_{\text{mix}} (\w)}{\eps^2 \, P_\text{in} ~ S_{b_1}^{(\text{mech})} (\w)} \sim \left( \frac{\eta_\text{mix}}{\eps} \right)^2 \sim \left( \frac{\delta_s}{\epsilon} \right)^2
\, ,
\ee
where we have normalized by the mechanical contribution to leakage noise in \Eq{Sdrive}. 

Both $\delta_s$ and $\epsilon$ parametrize the ability to control the geometry of the experiment and hence are treated together in Ref.~\cite{Bernard:2002ci}. Specifically, $\epsilon$ reflects the precision to which the loading and readout waveguide modes can be matched to the pump and signal modes, while $\delta_s$ reflects the precision to which the pump and signal modes can be matched to the ideal cylindrical ones. Both would be monitored and controlled by appropriate active feedback mechanisms. Thus, it is reasonable to estimate $\delta_s \sim \epsilon$ in the worst case, making mode mixing merely an $\order{1}$ correction to our existing treatment of mechanical leakage noise. In fact, since the cavity is larger than the waveguides, it would likely be possible to control it to a greater relative precision, $\delta_s \lesssim \epsilon$, in which case mode mixing is negligible. We thus do not include it in our sensitivity projections.

\subsection*{Cavity Frequency Drift}

In the previous section, we discussed how low frequency deformations of the cavity can lead to mixing between modes in the presence of higher frequency wall vibrations. Low frequency deformations alone do not lead to significant mode mixing, because the field in each mode adiabatically follows its slowly changing spatial profile. However, they can significantly affect the mode frequencies and the cross-coupling $\epsilon$. In the main body, we have addressed how $\epsilon$ must be actively monitored and controlled, as was already done in the MAGO experiment. In this section, we focus on the effect of mode frequency drift, which must be controlled similarly. 

Frequency drift manifests as an additional contribution to $\Delta(t)$ in \Eq{KirchoffMod}, which we write as $\Delta_s(t)$ in analogy to the slow deformations of the cavity walls $\delta_s(t)$. Unlike the elastic deformations considered for mechanical leakage noise, $\Delta_s(t)$ cannot be estimated from first principles, because it depends on technical details such as the cavity's hysteresis upon thermal expansion and contraction. However, since we are assuming the signal and pump modes can be held degenerate within their bandwidth, the RMS of the drift is bounded by
\be
\label{eq:DeltaRMS}
\Delta_s^{\text{rms}} = \frac{1}{2 \pi} \left(\int d\w ~ S_{\Delta_s}(\w)\right)^{1/2} \lesssim \frac{1}{Q_i} 
~.
\ee
The effect of cavity frequency drift is maximized if $S_{\Delta_s}(\w)$ is entirely supported at $|\w| \ll \Delta \w_d, \Delta \w_r$, in which case the integral in \Eq{SVR2} can be performed to give
\be
\label{eq:simple_SVi_expansion}
S_{V_i} (\w) \simeq \frac{(\w \, \w_i / Q_i)^2}{(\w^2 - \w_i^2)^2 + (\w \, \w_i / Q_i)^2}  ~ S_{V_d}(\w) \, \left(1 + (Q_i \, \Delta_s^{\text{rms}})^2 \right)
~.
\ee
Thus, perturbation theory breaks down entirely if \Eq{DeltaRMS} is no longer satisfied. In this case, however, we can still understand the effect of $\Delta_s$ on physical grounds: since the frequency drift is slow, the oscillations of the modes adiabatically follow it, implying that the pump mode and signal power will be spread over the frequency width $\Delta_s^{\text{rms}} \, \w_i$. This can be shown more precisely using the WKB approximation.\footnote{For the special case of ``monochromatic'' frequency wobble of amplitude $\delta \omega$ and frequency $\omega_{\Delta}$, where $V_i(t) \sim \exp(i \omega_i t) \exp(i (\delta \omega / \omega_{\Delta}) \cos(\omega_{\Delta} t))$, this can also be shown exactly using the Jacobi--Anger expansion. The $J_{n}(\delta \omega / \omega_{\Delta}) e^{i (\w_i + n \w_\Delta) t}$ terms have most of their weight for $|n| \sim \delta \w / \w_\Delta$, leading to the expected frequency spread of $\delta \w$.} Therefore, in the worst case scenario if $\Delta_s^\text{rms} \simeq 1/Q_i$, the power will at most be spread over the resonator width $\Delta \w_r$. This can be mimicked by replacing $\Delta \w_d \to \Delta \w_r$, as this also spreads out the pump mode and signal over frequency; we show the effect of this on the reach in the upper curve of the lower-right panel of Fig.~\ref{fig:reachvary}. 

We emphasize that as long as the pump and signal modes can be held degenerate, this is a maximally pessimistic assumption. First, $S_{\Delta_s}(\w)$ may have some of its support at frequencies $|\w| \gtrsim \Delta \w_r$, leading to an off-resonance suppression. For instance, if $S_{\Delta_s}(\w)$ is flat up to frequency $\Delta \w_s \gg \Delta \w_r$, then perturbation theory does not break down instead until $\Delta_s^{\text{rms}} \gtrsim \sqrt{\Delta \w_r / \Delta \w_s} / Q_i$. Furthermore, if $\Delta_s(t)$ is directly measured by the active feedback system that stabilizes the modes, it can be ``deconvolved'' almost entirely from the signal. As long as this can be done to a frequency precision of at least $\Delta \w_d$, low frequency noise does not affect the estimated reach. 

Again, we may compare this situation to that faced by interferometers, whose physical dimensions also drift. The fundamental reason that one can monitor the mode frequencies in our approach and subtract out its variations, but not do the same for an interferometer, is that typically the interferometer itself is the most sensitive ruler in the experiment. In our setup, one needs to only measure the signal and pump mode frequencies to fractional precision $\Delta \w_d / \w_0$, and atomic clocks exceed this by many orders of magnitude. 

As mentioned in the main body, the DarkSRF collaboration has already demonstrated frequency stabilization near that required by our most aggressive parameters. In addition, experimental tests of Lorentz invariance have stabilized cryogenic sapphire microwave oscillators to substantially greater precisions for $\order{\text{month}}$ timescales~\cite{Wolf:2002ip,Stanwix:2005yv,Stanwix:2006jb}. For our approach, even if a continuous run of length $\tint$ is infeasible, e.g.~if the cavity must be periodically recalibrated, an equivalent sensitivity can be attained by stitching together many shorter runs. Similarly, rare transient events that disrupt the experiment can be removed from the data stream.

\subsection*{Thermal and Amplifier Noise}

We adopt the same conventions as in Ref.~\cite{Berlin:2019ahk} to describe noise arising from thermal fluctuations of the cavity modes and the quantum-limited amplifier in the readout. For completeness, we derive the thermal noise PSD for the signal mode by applying the equipartition theorem to the equivalent RLC circuit. Thermal fluctuations of the signal mode can be modeled as sourced by the resistor, which drives the entire circuit with voltage $V_\text{th}$. Since the PSD of this noisy driving voltage is flat within the resonance width, we apply the narrow-width approximation to \Eq{SVR}, giving
\be
S_{V_1} \simeq \frac{\pi \w_1}{2 Q_1} \, \big( \delta (\w - \w_1) + \delta (\w + \w_1) \big) \, S_{V_\text{th}} (\w)
~.
\ee
Integrating over $\w$ thus leads to an average voltage across the resistor $R_1$ of
\be
\label{eq:thermal0}
\langle V_1^2 \rangle \simeq \frac{\w_1}{4 \pi \, Q_1} \, S_{V_\text{th}} (\w_1) ~.
\ee
By the equipartition theorem, the temperature of the circuit $T$ can be related to the energy stored in the inductor $L_1$, $T/2 \simeq L_1 \, \langle I^2 \rangle /2$, where $I$ is the current in the circuit. Since the voltage across the equivalent resistor of the signal mode is $V_1 = I \, R_1$, we have $\langle V_1^2 \rangle \simeq T \, R_1^2 / L_1$. Equating this to \Eq{thermal0} and using $Q_i = \w_i L_i / R_i$, we find $S_{V_\text{th}} \simeq 4 \pi T \, R_1$. 

However, only part of the resistance $R_1$ is due to the intrinsic dissipation $R_{\text{int}}$ of the circuit, and only this part necessarily sources thermal fluctuations. If the signal readout is connected to a cold load, so that it does not send thermal noise back to the cavity, then for the signal mode we actually have $S_{V_\text{th}} \simeq 4 \pi T \, R_{\text{int}}$. \Eq{SVR} then implies that the thermal noise PSD is
\be
\label{eq:thermal}
S_\text{th} (\w) = \frac{S_{V_1}(\omega)}{R_1} = \frac{Q_1}{Q_{\text{int}}} ~ \frac{4\pi T \, (\w \, \w_0 / Q_1)^2}{(\w^2 - \w_1^2)^2 + (\w \, \w_0 / Q_1)^2} 
~,
\ee
where we used $R_\text{int} / R_1 = Q_1 / Q_\text{int}$. 

The readout waveguide is attached to an amplifier, which sources its own noise. The lower bound on such noise is dictated by the standard quantum limit, arising from zero-point fluctuations and backaction/imprecision noise. The corresponding PSD is spectrally flat~\cite{RevModPhys.82.1155},
\be
\label{eq:amp}
S_\text{amp} (\w) = \pi \, \w_1 
\, .
\ee
We assume that amplifier noise is quantum-limited, which has been achieved in resonant cavity setups~\cite{Brubaker:2016ktl} and is often assumed for future projections of other axion experiments, such as DM Radio~\cite{Chaudhuri:2019ntz}. 

\section*{Expected Sensitivity}

\subsection*{Coupling Optimization}

Overcoupling the cavity to the readout corresponds to $Q_\text{cpl} \ll Q_\text{int}$. As discussed in Refs.~\cite{Chaudhuri:2018rqn,Chaudhuri:2019ntz,Berlin:2019ahk}, this is optimal for thermal noise limited resonant experiments, even though critical coupling maximizes the signal power, because it decreases \textit{both} the signal power and thermal noise in a way that allows a parametrically faster scan rate. Although these considerations do not apply to our broadband setup, it also benefits from overcoupling for the much simpler reason that it prevents an off-resonance signal from being overwhelmed by amplifier noise. In the limit where amplifier noise dominates, $Q_{\text{cpl}}$ should be as small as possible. 

For completeness, we now precisely describe how the signal and noise are affected by the value of $Q_{\text{cpl}}$. Recall that in the RLC analogy, the signal mode circuit has a resistor $R_1 = R_{\text{int}} + R_{\text{cpl}}$. When we computed the signal PSD, the thermal noise PSD, and the part of the leakage noise $S_{\text{leak}}^{(1)}$ corresponding to the loading waveguide coupling to the signal mode, we computed the total power dissipated across \textit{both} resistors. Thus, the fraction of power sent to the readout is smaller by a factor of $R_{\text{cpl}} / R_1 = Q_1 / Q_{\text{cpl}}$, and all of these PSDs should be rescaled by this amount. Amplifier noise is not affected, since it is intrinsic to the amplifier itself. Finally, consider the part of the leakage noise $S_{\text{leak}}^{(0)}$ corresponding to the readout waveguide coupling to the pump mode. In the RLC analogy, the pump mode circuit has a resistor $R_0 = R_{\text{int}} + \order{\epsilon^2} \, R_{\text{cpl}}$. Therefore, the fraction of power read out as leakage noise is proportional to $R_{\text{cpl}} / R_{\text{int}} = Q_{\text{int}} / Q_{\text{cpl}}$, and $S_{\text{leak}}^{(0)}$ should be rescaled by this factor. 

As described in the main body, we do not consider loaded quality factors lower than $Q_1 \sim 10^5$. One might worry that such a strong coupling to the signal mode might degrade the quality factor of the pump mode. Thus, we impose as a constraint that the power loss in the pump mode due to the readout is negligible, $\epsilon^2 R_{\text{cpl}} \ll R_{\text{int}}$, which implies 
\be
\label{eq:QcplBound}
Q_\text{cpl} \gtrsim \text{max} \left[ \eps^2 \, Q_\text{int} , 10^5 \right]
\, .
\ee
The constraint $Q_\text{cpl} \gtrsim \eps^2 \, Q_\text{int}$ is unimportant for almost all parameters we consider, except for the most conservative ones in the top-right panel of \Fig{reachvary}. Critical coupling is optimal for the lowest axion masses, while overcoupling as much as possible is optimal for the highest axion masses. For each intermediate axion mass, a different intermediate coupling is optimal, because overcoupling increases the strength of leakage noise. However, we find numerically that essentially all of the reach shown in Fig.~\ref{fig:reach} can be obtained using only a critically coupled run and a maximally overcoupled run. A small remaining slice of parameter space at small couplings and intermediate axion masses can be covered using a third run with $Q_\text{cpl} \sim 10^7$.

\subsection*{Statistics for Expected Exclusion}

In this section, we roughly describe the statistics of a broadband low mass axion search, with the main purpose of explaining why the expected sensitivity decreases for $\tint \lesssim \tau_a$, where $\tau_a \sim Q_a / m_a$ is the axion coherence time. A Bayesian approach to the same problem is given in Ref.~\cite{Centers:2019dyn}. A similar frequentist approach is given in Ref.~\cite{Foster:2017hbq}, though it focuses on the case $\tint \gg \tau_a$. 

For concreteness, we neglect unvirialized components of the axion field, as well as any enhanced structure in the axion field that could arise, e.g., from strong axion self-interactions or parametric resonance effects~\cite{Arvanitaki:2019rax, Fairbairn:2017sil, Kolb:1993zz, Kolb:1993hw, Kolb:1994fi, Kolb:1995bu}. In the absence of such effects, in the Milky Way the axion can be described as a collection of classical plane waves with independent phases. An experiment with total integration time $\tint$ can only resolve frequency bins of width $\Delta \w \sim 1/\tint$. Each bin contains macroscopically many axions; for instance, for $\tint \gtrsim \tau_a$, 
\be
\Delta N_a \simeq \frac{\rhodm V_\text{cav}}{m_a} \, \frac{\tau_a}{\tint} \simeq 10^{25} \left(\frac{10^{-14}\;{\rm eV}}{m_a}\right)^2\left(\frac{5\;{\rm years}}{\tint}\right)\left(\frac{V_{\text{cav}}}{\rm m^3}\right)
~,
\ee
so the central limit theorem applies to the amplitude in each bin. Specifically, suppose we measure $a(t)$ for a time $\tint$ and perform a discrete Fourier transform (DFT), yielding the complex amplitude $\tilde{a}_i$ for the frequency bin centered at $\w_i$. Then the real and imaginary parts of $\tilde{a}_i$ are independent Gaussian random variables with zero mean,\footnote{Under the standard DFT, the amplitudes $\tilde{a}_i$ in neighboring bins will actually be slightly correlated. We neglect this small effect below.} so the axion field can be treated as a Gaussian random field. For the rest of this section we will use a PSD normalization suited for these DFT elements, rather than the continuous normalization of \Eq{PSDdef}. For example, for the axion field we define
\be
\langle \tilde{a}_{i} \tilde{a}^*_{j} \rangle = \delta_{ij} \, S_a(\w_i) \label{eq:axion_PSD}
~,
\ee
where $S_a (\w_i)$ is the \emph{discrete} PSD. As illustrated in Ref.~\cite{Centers:2019dyn}, a typical realization of $a(t)$ is approximately monochromatic on timescales $\tint \lesssim \tau_a$, but fluctuates in amplitude on timescales $\tau_a$ with respect to the RMS value $\sqrt{\langle a(t)^2\rangle} = \sqrt{\rhodm}/m_a$. For $\tint \lesssim \tau_a$, the amplitude is approximately fixed for the duration of the experiment, and the possibility of observing a downward amplitude fluctuation is responsible for weakening the projected sensitivity.

For simplicity, we will specialize to axion detection experiments using static fields, and return to our heterodyne approach later. For a static field experiment, the frequency components of the signal $s(t)$ are simply those of the axion field multiplied by a frequency-dependent filtering. Therefore, the signal can also be treated as a Gaussian random field. The experiment measures a data stream $d(t) = s(t) + n(t)$, where the noise $n(t)$ is independent of the signal. For the noise sources that we consider, $n(t)$ is also a Gaussian stationary random variable with zero mean. Thus, the likelihood of observing the data is
\be
\label{eq:data_likelihood}
L[\tilde{d}] = \prod_i  \, \frac{e^{- |\tilde{d}_i|^2 / (S_s(\omega_i) + S_n(\omega_i))}}{\pi (S_s(\omega_i) + S_n(\omega_i))}
\ee
where the frequency bins have width $\Delta \w = 2 \pi / t_{\text{int}}$. We note that this result has been previously derived in Ref.~\cite{Foster:2017hbq}. 

We assume for simplicity of notation that the data is taken in a single continuous run, but this is not necessary, as distinct runs can be stitched together. In fact, given a fixed integration time $t_{\text{int}}$, this can actually be advantageous. As long as $t_{\text{exp}} \gg \tau_a$, where $t_{\text{exp}}$ is the total duration of the experiment, the reach will not be penalized by the effect discussed above because the distinct runs during the experiment will sample different amplitudes for the axion field. In addition, \Eq{data_likelihood} implicitly assumes that the axion oscillates many times during the experiment, $t_{\text{exp}} \gg 1/m_a$. For $t_{\text{exp}} \lesssim 1/m_a$, the likelihood additionally depends on the instantaneous phase of the axion field, which leads to an additional $\order{1}$ suppression of the reach; we will not consider this case below.

The average signal and noise PSDs $S_s(\w_i)$ and $S_n(\w_i)$ also depend on nuisance parameters $\theta_{s, n}$ that we imagine are measured with calibration runs. For the purposes of placing an exclusion on $\gyy$, it is convenient to define 
\be
\gyy^2 \, \lambda_{s, i}(\theta_s) \equiv S_s(\w_i, \theta_s) \, , \quad \lambda_{n, i}(\theta_n) \equiv S_n(\w_i, \theta_n)
\ee
so that the likelihood takes the form
\be
\label{eq:Lik1}
L(\gyy, \theta_s, \theta_n) = \prod_i \, \frac{e^{- |\tilde d_i|^2/ (\gyy^2 \lambda_{s, i}(\theta_s) + \lambda_{n, i}(\theta_n))}}{\pi (\gyy^2 \, \lambda_{s, i}(\theta_s) + \lambda_{n, i}(\theta_n))} ~ L_{\rm aux}(\theta_s, \theta_n) \, ,
\ee
where $L_{\rm aux}$ contains the results of calibration measurements and is not necessarily Gaussian. These measurements are independent of the data that we take during our physics run, so the two probabilities multiply. 

Let $\ghyy$ be the maximum likelihood estimator for $\gyy$. The incompatibility of the coupling value $\gyy$ with the data can be quantified by the test statistic~\cite{Cowan:2010js}
\be
\label{eq:TS1}
q(\gyy) = -2 \, \log{\left(\frac{L(\gyy, \hat{\hat\theta}_s, \hat{\hat\theta}_n)}{L(\ghyy, \hat{\theta}_s, \hat{\theta}_n)} \right)} ~ \Theta (\gyy^2 - \ghyy^2)\, ,
\ee
where $\ghyy$ and $\hat{\theta}_{s,n}$ are unconditional maximum-likelihood estimators and $\hat{\hat{\theta}}_{s,n}$ are conditional maximum-likelihood estimators for fixed $\gyy$. The step function reflects the fact that we should not be able to exclude couplings smaller than the best-fit value. Below, we will assume $L_{\rm aux}(\hat \theta_s, \hat \theta_n)\simeq L_{\rm aux}(\hat{\hat \theta}_s,\hat{\hat \theta}_n)$, so the nuisance parameters play little role. 

When the integration time is much longer than the axion coherence time, $\tint \gg \tau_a$, the axion signal is spread over many bins, and asymptotic theorems apply. In particular, Wilks' theorem~\cite{Wilks:1938dza} implies that the distribution of $q(\gyy)$ for fixed $\gyy$ is a half chi-squared distribution with one degree of freedom, implying that the 90\% and 95\% C.L. upper bounds are
\be
\label{eq:qcuts}
q_{90\%} = 1.64, \quad q_{95\%} = 2.71.
\ee
Assuming that no axion exists, the exclusion that can be set varies from trial to trial. We use the approach illustrated in Ref.~\cite{Cowan:2010js}, where it is shown that the median exclusion is achieved by the so-called Asimov dataset, in which each of the $|\tilde{d}_i|^2$ are set to the mean value achieved in a background-only dataset, i.e., $|\tilde{d}_i|^2 \to \lambda_{n, i}$. In this case, $\ghyy^2 = 0$. Using this in \Eq{TS1} and approximating $\gyy^2 \, \lambda_{s, i} \ll \lambda_{n, i}$ for all $\gyy$ near the sensitivity threshold (valid because the signal is spread over many bins) gives 
\be
q(\gyy) \simeq \sum_i \left( \frac{\gyy^2 \, \lambda_{s, i}(\hat{\hat \theta}_s)}{\lambda_{n, i}(\hat{\hat \theta}_n)} \right)^2 ~,
\ee
where we assumed $\lambda_n(\hat \theta_n)\simeq \lambda_n(\hat{\hat \theta}_n)$. This is closely related to the signal-to-noise ratio (SNR) used to estimate the reach in many axion experiments (see, e.g., Ref.~\cite{Berlin:2019ahk}), as can be shown by approximating the sum in the above expression as an integral,
\be
\label{eq:SNR}
q(\gyy) \simeq \frac{\tint}{2 \pi} \int_0^\infty d \omega \, \left( \frac{S_s (\w)}{S_n (\w)} \right)^2 = \text{SNR}^2 \, .
\ee
Since this result involves a ratio of PSDs, it also holds for the continuous PSD normalization of \Eq{PSDdef}. Here, negative frequency bins were not included since they are not independent of the positive frequency bins. Combining this with \Eq{qcuts} implies that the median 90\% or 95\% expected exclusion corresponds to an SNR of 
\be
 \label{eq:SNR1}
\text{SNR} (\tint \gg \tau_a) \gtrsim
\begin{cases}
1.3 & 90\% \text{ C.L.}\\
1.6 & 95\% \text{ C.L.}
~,
\end{cases}
\ee
which roughly matches the $\text{SNR} \gtrsim 1$ prescription commonly adopted in the axion literature (see, e.g., Refs.~\cite{Chaudhuri:2014dla,Kahn:2016aff}). 

In the short integration time limit $\tint \ll \tau_a$, the axion signal cannot be resolved, and hence lies in a single frequency bin.\footnote{More precisely, the axion signal could straddle two frequency bins; we neglect this small effect.} In the following we omit for simplicity the explicit dependence on nuisance parameters and $L_{\rm aux}$. We continue to assume negligible systematic errors: $\lambda_n(\hat \theta_n)\simeq \lambda_n(\hat{\hat \theta}_n)$ and $L_{\rm aux}(\hat \theta_s, \hat \theta_n)\simeq L_{\rm aux}(\hat{\hat \theta}_s,\hat{\hat \theta}_n)$. Dropping the $i$ subscript and defining $S \equiv |\tilde{d}_i|^2$, we have
\be
\label{eq:Lik2}
L(\gyy^2) = \frac{e^{-S/(\gyy^2 \, \lambda_s + \lambda_n)}}{\gyy^2 \, \lambda_s + \lambda_n} 
~.
\ee
In this case, Wilks' theorem does not apply, but the calculation of the test statistic is analytically tractable. In particular, $\ghyy^2$ can be found by analytically maximizing the likelihood, giving
\be
\ghyy^2 = \begin{cases} (S - \lambda_n) / \lambda_s & S \geq \lambda_n \\ 0 & S < \lambda_n \end{cases}
\ee
where the second line is a consequence of $\gyy^2 > 0$. The test statistic then takes the explicit form
\be
\label{eq:TS2}
q(\gyy^2, S) = 2 \times
\begin{cases}
0 & \gyy^2 \, \lambda_s + \lambda_n < S
\\
\frac{S}{\gyy^2 \, \lambda_s + \lambda_n} - 1 + \log{\frac{\gyy^2 \, \lambda_s + \lambda_n}{S}} & \lambda_n \leq S \leq \gyy^2 \, \lambda_s + \lambda_n
\\
\frac{S}{\gyy^2 \, \lambda_s + \lambda_n} - \frac{S}{\lambda_n} + \log{\frac{\gyy^2 \, \lambda_s + \lambda_n}{\lambda_n}} & S < \lambda_n
~.
\end{cases}
\ee
At fixed $\gyy^2$, $q (\gyy^2, S)$ is a monotonically decreasing function of $S$. Thus, to compute the upper bound $q_\alpha$ on $q$ corresponding to a given C.L. $\alpha$, we can find the value $S_\alpha$ such that the probability for $S \leq S_\alpha$ is $P(S\leq S_\alpha)=1-\alpha$ in order to obtain $q_\alpha(\gyy^2)=q (\gyy^2, S_\alpha(\gyy^2))$. Using the known distribution of $S$ in \Eq{Lik2} for a given axion coupling $\gyy^2$, we have
\be
\int_0^{S_\alpha(\gyy^2)} dS ~ \frac{e^{-S / (\gyy^2 \, \lambda_s + \lambda_n)}}{\gyy^{2} \lambda_s + \lambda_n} = 1 - \alpha
~.
\ee
Solving for $S_\alpha$ then yields
\be
S_\alpha (\gyy^2) = \left|\log \alpha \right| (\gyy^2 \, \lambda_s + \lambda_n)
~.
\ee
Therefore, in the event that there is no axion signal, the median expected exclusion for an experiment at $(100 \times \alpha) \%$ C.L. is determined by solving~\cite{Cowan:2010js}
\be
q_\alpha(\gyy^2) = q \big(\gyy^2 , S_{0.5} (0) \big)
~.
\ee
Once again identifying $q(\gyy^2) \simeq \text{SNR}^2$, we find that the median expected 90\% or 95\% limit on $\gyy$ corresponds to 
\be
 \label{eq:SNR2}
\text{SNR} (\tint \ll \tau_a) \gtrsim
\begin{cases}
5.6 & 90\% \text{ C.L.} \\
12.5 & 95\% \text{ C.L.}
\end{cases}
\ee
Since the SNR is proportional to $\gyy^2$, the higher threshold in \Eq{SNR2} compared to \Eq{SNR1} corresponds to weakening the 90\%--95\% C.L. sensitivity projections for $\gyy$ by a factor of $2$--$3$ when $\tint \lesssim \tau_a$. 

For comparison, Ref.~\cite{Centers:2019dyn} instead found a weakening of $\sim 4$ for the 95\% C.L. $\gyy$ projections, using a Monte Carlo estimate for the test statistic sampling distribution. That work also found a weakening factor of $\sim 10$ at 95\% C.L.~using a Bayesian approach with a flat prior on $\gyy$. However, a flat prior in $\log \gyy$ is also reasonable on subjective grounds, as evidenced by the common use of logarithmic scales in plots like Fig.~\ref{fig:reach}. The logarithmic prior penalizes smaller values of $\gyy$ much less, and thus the weakening of the sensitivity for $\tint \lesssim \tau_a$ is more mild. Similarly, a flat prior in $\gyy^2$ would also be reasonable since the signal is proportional to it, but this penalizes smaller values of $\gyy$ to a greater degree, enhancing the sensitivity suppression. Since the conclusions of the Bayesian approach vary significantly between reasonable priors, we adopt the frequentist approach described above. 

In the above analysis, we have mainly sought to explain analytically why the reach is weakened for $\tint \lesssim \tau_a$. Since this effect arises solely from the fluctuations of the axion field amplitude, we expect that a similar penalty factor should apply for our heterodyne approach. However, showing this analytically would be notationally complex, because the axion Fourier components are spread out by, e.g., the width of the driver $\Delta \w_d$, which simultaneously affects the noise. Thus, we defer a more detailed numerical calculation of the projected sensitivity to future work. To estimate our reach here, we use Eqs.~(\ref{eq:SNR}), (\ref{eq:SNR1}), and (\ref{eq:SNR2}), along with the following small modification: for a static field experiment, bins at positive and negative frequencies $\pm \omega$ are redundant because the data stream is real-valued. For a heterodyne experiment, bins at $\w_0 \pm \w$ are redundant for the same reason, so \Eq{SNR} should only integrate over positive frequencies above $\w_0$. The sole exception is when amplifier noise dominates, since its contributions at frequencies $\w_0 \pm m_a$ are independent of each other.

\newpage
\section*{Table of Notation}

\renewcommand{\arraystretch}{1.4}
\begin{table}[h!]
\centering
\begin{tabular}{|c||c|c|}
\hline
\textbf{Symbol} & \textbf{Meaning} & \textbf{Reference} \\
\hline\hline
$B_i$ & volume-averaged pump ($i = 0$), signal ($i=1$) mode magnetic field & \Eq{B0} \\ \hline
$\Evec_i(x)$, $\Bvec_i(x)$ & spatial profile of the pump ($i = 0$), signal ($i=1$) mode fields & \Eq{formfactor} \\ \hline
$M_\text{cav}$ & mass of the cavity & $-$ \\ \hline
$P_\text{in}$ & power stored in the pump mode & \Eq{Pin} \\ \hline
$q_\text{rms}$ & attenuated RMS displacement of cavity walls & \Eq{Sfm} \\ \hline
$Q_a$ & effective axion quality factor & pg.~\pageref{eq:photon} \\ \hline
$Q_\text{int}$ & intrinsic quality factor of cavity & \Eq{Qtot} \\ \hline
$Q_\text{cpl}$ & coupling to the readout & Eqs.~(\ref{eq:Qtot}), (\ref{eq:QcplBound}) \\ \hline
$Q_i$ & quality factor of the pump ($i = 0$), signal ($i=1$) mode & pg.~\pageref{eq:Qtot} \\ \hline
$Q_m$ & quality factor of cavity mechanical resonance & pg.~\pageref{eq:elastic_deformation_eq} \\ \hline
$S_a$ & PSD of the axion field& \Eq{Sa} \\ \hline
$S_\text{amp}$ & PSD of the readout amplifier & \Eq{amp} \\ \hline
$S_{b_i}$ & normalized PSD of the pump ($i = 0$), signal ($i=1$) mode & \Eq{Sbidef} \\ \hline
$S_{b_i}^{\text{(phase)}}$ & additive correction to $S_{b_i}$ from oscillator phase noise & \Eq{Sbi} \\ \hline
$S_{b_i}^{\text{(mech)}}$ & additive correction to $S_{b_i}$ from mechanical vibrations & \Eq{SbiMech2} \\ \hline
$S_\text{sig}$ & PSD of axion signal & \Eq{SsigApprox} \\ \hline
$S_\text{th}$ & PSD of thermal noise & \Eq{thermal} \\ \hline
$S_\ph$ & PSD of oscillator phase noise & \Eq{cphase} \\ \hline
SNR & signal-to-noise ratio & \Eq{SNR} \\ \hline
$\tint$ & experimental integration time & $-$ \\ \hline
$V_\text{cav}$ & volume of the cavity & $-$ \\ \hline
$\delta$ & fractional displacement of cavity walls & \Eq{SbiMech2} \\ \hline
$\eps$ & suppression of leakage noise & \Eq{Sdrive} \\ \hline
$\eta_a$ & form factor of axion signal & \Eq{formfactor} \\ \hline
$\eta_\text{mix}$ & form factor of mechanical mode mixing & \Eq{etamix} \\ \hline
$\w_i$ & frequency of the pump ($i = 0$), signal ($i=1$) mode & pg.~\pageref{eq:photon} \\ \hline
$\w_\text{sig}$ & frequency of the axion signal & \Eq{wsig} \\ \hline
$\w_m$ & frequency of a cavity mechanical resonance & \Eq{wmech} \\ \hline
$\w_\text{min}$ & lowest-lying mechanical resonance & \Eq{Sfm} \\ \hline
$\Delta \w_a$ & width of the axion field & pg.~\pageref{eq:photon} \\ \hline
$\Delta \w_d$ & width of the external driving oscillator & \Eq{DeltaDriver} \\ \hline
$\Delta \w_r$ & width of the cavity resonance & pg.~\pageref{eq:Psig1} \\ \hline
$\Delta \w_\text{sig}$ & signal bandwidth & \Eq{SNR0} \\ \hline
\end{tabular}
\end{table}

\end{document}